\documentclass[sigconf]{acmart}

\AtBeginDocument{%
	\providecommand\BibTeX{{%
			Bib\TeX}}}

\copyrightyear{2024}
\acmYear{2024}
\setcopyright{acmlicensed}
\acmConference[AsiaCCS'24]{Proceedings of the 2024 ACM ASIA Conference on Computer and Communications Security}{July 1--5, 2024}{Singapore}
\acmBooktitle{Proceedings of the 2024 ACM ASIA Conference on Computer and Communications Security (AsiaCCS '24), July 1--5, 2024, Singapore}
\acmDOI{10.1145/3634737.3637644}
\acmISBN{979-8-4007-0482-6/24/07}

\usepackage[normalem]{ulem}

\usepackage{algorithmic}
\usepackage{graphicx}
\usepackage{textcomp}
\usepackage{xcolor}
\usepackage{graphicx,float,wrapfig}
\usepackage{color,soul}
\usepackage{xspace}
\usepackage[utf8]{inputenc}
\usepackage{listings}
\usepackage{xcolor}
\usepackage{xurl}
\usepackage{hyperref}
\usepackage{pifont}
\usepackage{array}
\usepackage{comment}
\usepackage{arydshln}
\usepackage{circledsteps}
\usepackage{fmtcount}
\usepackage{subfig}
\usepackage{multirow}
\usepackage{threeparttable}
\usepackage{array}
\usepackage{enumitem}
\usepackage{cprotect}
\newcommand{\cmark}{\ding{51}}%
\newcommand{\xmark}{\ding{55}}%

\definecolor{codegreen}{rgb}{0,0.6,0}
\definecolor{codegray}{rgb}{0.5,0.5,0.5}
\definecolor{codepurple}{rgb}{0.58,0,0.82}
\definecolor{backcolour}{rgb}{0.95,0.95,0.92}

\pgfkeys{/csteps/inner xsep=1.5pt}
\pgfkeys{/csteps/inner ysep=1.5pt}
\pgfkeys{/csteps/inner color=white} 
\pgfkeys{/csteps/outer color=black} 
\pgfkeys{/csteps/fill color=black}  

\colorlet{punct}{red!60!black}
\definecolor{background}{HTML}{EEEEEE}
\definecolor{delim}{RGB}{20,105,176}
\colorlet{numb}{magenta!60!black}

\lstdefinestyle{mystyle}{
    backgroundcolor=\color{backcolour},   
    commentstyle=\color{codegreen},
    keywordstyle=\color{magenta},
    numberstyle=\tiny\color{codegray},
    stringstyle=\color{codepurple},
    basicstyle=\ttfamily\footnotesize,
    breakatwhitespace=false,         
    breaklines=true,                 
    captionpos=b,                    
    keepspaces=true,                 
    numbers=left,                    
    numbersep=5pt,                  
    showspaces=false,                
    showstringspaces=false,
    showtabs=false,                  
    tabsize=2,
    literate=
     *{0}{{{\color{numb}0}}}{1}
      {1}{{{\color{numb}1}}}{1}
      {2}{{{\color{numb}2}}}{1}
      {3}{{{\color{numb}3}}}{1}
      {4}{{{\color{numb}4}}}{1}
      {5}{{{\color{numb}5}}}{1}
      {6}{{{\color{numb}6}}}{1}
      {7}{{{\color{numb}7}}}{1}
      {8}{{{\color{numb}8}}}{1}
      {9}{{{\color{numb}9}}}{1}
      {:}{{{\color{punct}{:}}}}{1}
      {,}{{{\color{punct}{,}}}}{1}
      {\{}{{{\color{delim}{\{}}}}{1}
      {\}}{{{\color{delim}{\}}}}}{1}
      {[}{{{\color{delim}{[}}}}{1}
      {]}{{{\color{delim}{]}}}}{1}     
}

\def\BibTeX{{\rm B\kern-.05em{\sc i\kern-.025em b}\kern-.08em
    T\kern-.1667em\lower.7ex\hbox{E}\kern-.125emX}}
    
\newcommand{\sysname}{{\scshape BYOTee}\xspace}
\newcommand{\firmware}{{\scshape Firmware}\xspace}
\newcommand{\hbuilder}{{\scshape HardwareBuilder}\xspace}
\newcommand{\SSAPacker}{{\scshape SSAPacker}\xspace}

\newcommand{\modela}{{\textsf{BaseModel}}\xspace}
\newcommand{\modelb}{{\textsf{EnhancedModel}}\xspace}
\newcommand{\hwloader}{{\scshape Hw-Att}\xspace}

\newcounter{goal}
\newcommand{\goal}[1]{
  \stepcounter{goal}
  \emph{G\padzeroes[1]\decimal{goal}. #1}
}

\title{Building Your Own Trusted Execution Environments Using FPGA}

    \author{Md Armanuzzaman}
    \affiliation{%
  	\institution{CactiLab, University at Buffalo}
  	\city{Buffalo}
  	\country{USA}}
	\email{mdarmanu@buffalo.edu}
    
    \author{Ahmad-Reza Sadeghi}
	\affiliation{%
  	\institution{Technische Universität Darmstadt}
  	\city{Darmstadt}
  	\country{Germany}}
	\email{ahmad.sadeghi@trust.informatik.tu-darmstadt.de}
	        
    \author{Ziming Zhao}
	\affiliation{%
  	\institution{CactiLab, University at Buffalo}
  	\city{Buffalo}
  	\country{USA}}
	\email{zimingzh@buffalo.edu}

\begin{document}

\begin{abstract}
Despite of their benefits, existing Trusted Execution Environments (TEE) or enclaves have been criticized for lack of transparency, vulnerabilities, and various restrictions. 
A significant limitation is that they only provide a static and fixed hardware Trusted Computing Base (TCB) that cannot be customized for different applications. 
The design violates the principle of least privilege by including unnecessary peripherals in the hardware TCB and buggy peripheral drivers in the software TCB.
Additionally, Existing TEEs time-share a processor core with the Rich Execution Environment (REE), making execution less efficient and vulnerable to cache side-channel attacks. 
Although many previous projects have focused on addressing software issues in TEEs on SGX, TrustZone, or RISC-V, some TEE issues are inherent in the hardware system's design, making them impossible to resolve with software alone.

In this paper, we present \sysname (\underline{B}uild \underline{Y}our \underline{O}wn \underline{T}rusted \underline{E}xecution \underline{E}nvironments), which is an easy-to-use hardware and software co-design infrastructure for building enclaves using Field Programmable Gate Arrays (FPGA).
\sysname creates enclaves with customized hardware TCBs and establishes a dynamic root of trust that allows untampered execution of Security-Sensitive Applications (SSA) from preexisting software on the hardcore system.
Additionally, \sysname provides mechanisms to attest the integrity of enclaves' hardware and software stacks. 
We implement a \sysname system for the Xilinx System-on-Chip (SoC) FPGA. 
The evaluations on the low-end Zynq-7000 system for four SSAs and 12 benchmark applications demonstrate the usage, security, effectiveness, and performance of the \sysname framework.
\end{abstract}



\maketitle


\section{Introduction}
\label{s:intro}

Existing Trusted Execution Environments (TEEs) on commodity computing devices rely on CPU hardware security primitives to ensure the confidentiality and integrity of code and data loaded within them, while protecting them from the Rich Execution Environment (REE). 
These hardware security primitives are provided by the CPU and can either be proprietary, as in Intel SGX~\cite{costan2016intel} and Arm TrustZone~\cite{pinto2019demystifying}, or open-sourced, as in RISC-V~\cite{bahmani2021cure}.
In recent years, we have witnessed unprecedented growth in using such TEEs in real-world products and academic projects,
which include real-time kernel protections~\cite{azab2014hypervision,ge2014sprobes}, securing containers and libraries~\cite{arnautov2016scone,tsai2017graphene, santos2014using}, shielding applications from attacks~\cite{baumann2015shielding,lind2017glamdring,schuster2015vc3}.
However, the hardware layer of current TEEs suffers from various issues, rendering them untrustworthy or ineffective. 



Firstly, they only offer a static and fixed hardware Trusted Computing Base (TCB) that cannot be customized for different applications at \emph{runtime}. 
Although RISC-V allows for hardware customization at design and manufacturing time, the resulting hardware configuration is static and cannot be changed after manufacturing.
The hardware primitives of TrustZone give the TEE the highest privilege to control the REE and communicate with all peripherals, thereby violating the principle of least privilege. 
It also includes unnecessary peripherals and buggy peripheral drivers in the software TCB~\cite{cerdeira2020sok}, exposing the TEE to malicious peripheral inputs~\cite{gross2019breaking}. 
Meanwhile, SGX's hardware requires applications in enclaves to trust the REE Operating System (OS) to communicate with peripherals~\cite{schneider2020pie}, bloating the size of the software TCB by including a usually monolithic REE OS kernel.

Another significant issue 
is that most commercially popular TEE hardware, such as SGX and TrustZone, are proprietary and limited to specific architectures, which require users to place blind trust in their security. 
As a result, users cannot verify the correctness of the TEE designs. 
Unfortunately, vulnerabilities, such as cache side-channels, have been discovered in both SGX and TrustZone~\cite{zhang2016truspy, cho2018prime, brasser2017software, zhang2016return, gutierrez2018cachelight}, which undermines their security promises. 
Additionally, the proprietary nature of such TEE systems poses a challenge for researchers to explore the security properties and capabilities of different TEE configurations.
Although open-sourced TEE designs based on RISC-V can be verified at the design stage, dynamically attesting hardware states at runtime remains an unsolved problem.

Software-based solutions alone on existing hardware TEEs cannot address the aforementioned issues, which are rooted in the design of their respective hardware systems.
For example, {\scshape Sanctuary}~\cite{brasser2019sanctuary} configures the memory access controller to provide multi-domain isolation for sensitive applications on TrustZone, 
and {\scshape Cure}~\cite{bahmani2021cure} enables the exclusive assignment of system resources, e.g., peripherals, CPU cores, or cache resources, to each enclave on RISC-V.
While these solutions are noteworthy, they do not provide a customizable and attestable hardware TCB at runtime.
Other hardware-based solutions, such as HECTOR-V~\cite{nasahl2020hector} and Graviton~\cite{volos2018graviton}, do not address these issues either, and they cannot be deployed on commodity devices due to the need for hardware modifications.

In this paper, we present a hardware and software co-design framework to \underline{B}uild \underline{Y}our \underline{O}wn \underline{T}rusted \underline{E}xecution \underline{E}nvironments (\sysname).
\sysname utilizes commodity System-on-Chip (SoC) Field Programmable Gate Arrays (FPGAs), e.g., AMD EPYC FPGA-infused CPU~\cite{AMDFPGA} and Xilinx SoC FPGA, without requiring any hardware changes.  
With the \sysname toolchain, users can quickly and easily build multiple secure and customized enclaves on-demand to execute their Security-Sensitive Applications (SSA).
Each enclave is designed to include only the hardware and software necessary for the SSA and excludes other hardware and software components on the system, minimizing the sizes of hardware and software TCB.  

A SoC FPGA system, including FPGA-infused CPUs, integrates both a hardcore CPU, e.g., x86/64, Cortex-A or RISC-V, and an FPGA programmable logic architectures. 
The nature of FPGA enables on-demand configurations of enclaves' hardware TCBs, which may include softcore CPUs, Block RAM based (BRAM; same as Static RAM on known SoC FPGA devices) main memory, and peripherals.
Each enclave in \sysname has its own isolated physical address space, which maps its own dedicated main memory, system configuration registers, and peripherals.
The software on the hardcore CPU and other enclaves cannot access an enclave's address space unless it is explicitly specified in the design. 
By assigning the hardware resources to co-resident enclaves, \sysname creates a multiprogramming environment, isolates software faults, and provides memory protection on FPGAs. 
Enclaves in \sysname do not share processors with each other or the hardcore system, which eliminates the cache side-channel attack vector. Additionally, due to the characteristics of BRAM, cold-boot attacks on these enclaves are challenging. 

\sysname utilizes the \emph{secure configuration} process of the FPGA to establish a \emph{dynamic root of trust} that ensures complete isolation and untampered execution of Security-Sensitive Applications (SSAs) in enclaves from preexisting software on the hardcore system, including the hypervisor and operating system. 
Additionally, \sysname offers both software- and hardware-based remote attestation mechanisms that operate under two threat models. 
To enable execution of SSAs, external libraries and drivers for peripherals are required. 
On the software front, the configurable \firmware component of \sysname provides essential software libraries such as libc, as well as a Hardware Abstraction Layer (HAL), to minimize the software Trusted Computing Base (TCB).



We have implemented the \sysname infrastructure and toolchain for the Xilinx SoC FPGA. 
The toolchain comprises several components, including \hbuilder, which takes developers' hardware resource requirements as input and generates hardware modules and interconnections. 
By automating this process, \hbuilder reduces the likelihood of developer-induced misconfigurations and allows developers to focus on SSA development, thereby increasing the usability of \sysname. 
Additionally, the system and toolchain include \hwloader, a trusted decryption and attestation hardware module implemented in bitstream, \firmware, a software runtime for SSAs, and \SSAPacker, a tool used to encrypt and sign an SSA binary.
The contributions of this paper are as follows:

\begin{itemize}[leftmargin=*]
\item We present \sysname, a framework to create enclaves with minimal hardware and software TCBs on commodity SoC FPGA. 
The framework can also help researchers who want to explore the capabilities and security properties of different TEE hardware configurations.
The idea of \sysname can be implemented on FPGA systems with a secure configuration process from any vendor;

\item \sysname establishes a dynamic root of trust that allows full isolation and untampered execution of SSAs in enclaves from preexisting software on the hardcore system;

\item We present software- and hardware-based remote attestations for two threat models to capture the identities of the enclave hardware configurations, firmware, and SSAs;

\item We implement the \sysname system and toolchain for the Xilinx SoC FPGA. 
    We open-source the \sysname system and toolchain\footnote{https://github.com/CactiLab/BYOTee-Build-Your-Own-TEEs};

\item We demonstrate \sysname's usage, security, effectiveness, and performance with the Embench-IoT benchmark and four SSAs on the low-end MicroBlaze softcore CPU and Zynq-7000 system.
\end{itemize}

\section{SoC FPGA and Root of Trust}
\label{s:back}

In this section, we provide an overview of FPGA and the hardware modules and root of trust commonly found on a SoC FPGA. 
We will also discuss the workflow involved in designing, developing, and securely configuring a SoC FPGA.

\subsection{Benefits of FPGA and FPGA in End Products} 
FPGA is designed to be configured by users using Register-Transfer Level (RTL) code after manufacturing.
Besides reconfigurability, it has advantages of high performance, fast development round, etc.
While FPGAs were mainly used for hardware prototyping years ago, their benefits and reduced costs have made them practical for end products recently~\cite{mscatapult, msbrainve}.
A variety of FPGA products ranging from embedded systems, i.e., Xilinx Zynq-7000 with only 3,600 logic elements ($\approx$\$70), to data center devices, i.e., Agilex F R25A with 2.6 millions of logic elements ($\approx$\$10k), are available.
Amazon Elastic Compute Cloud has been offering FPGAs to their customers in F1 instances since 2017~\cite{amazonF1}, 
and AMD starts infusing EPYC CPUs with FPGA in 2023~\cite{AMDFPGA}.
Various FPGA-based application-specific accelerators, such as deep neural networks~\cite{zhang2015optimizing, suda2016throughput, li2019rnn}, classic and post-quantum cryptographic algorithms~\cite{chelton2008fast, elkhatib2021high},  
Memcached~\cite{lavasani2013fpga}, 
have been proposed and deployed. 

FPGA can be used to build general-purpose computing platforms, in which users can design and implement their own softcore CPUs or customize existing open-sourced~\cite{OpenCores, VexRiscv, Neo430, microwatt, a2i, a2o, OpenSPARC, libreSoc} or proprietary ones~\cite{lysecky2009design, nios}.
The available softcore CPUs range from the partially configurable (e.g., cache size, pipeline depth)
and proprietary low-end 32-bit MicroBlaze~\cite{lysecky2009design}, to the fully customizable and open-sourced mid-end 32-bit RISC-V~\cite{VexRiscv, matthews2017taiga} and high-end 64-bit A2I POWER processor~\cite{a2i}.
Even though existing softcore CPUs on FPGA have a lower maximum clock frequency, i.e., 500MHz, than their hardcore counterparts, i.e., 4GHz, 
softcore and hardcore CPUs with similar complexity and frequency have comparable performance.
While the low-end softcore CPUs are comparable to microcontrollers, the mid-end and high-end ones have performances comparable to hardcore microprocessors~\cite{heinz2019catalog}.
Because it is possible to formally verify RTL implementations~\cite{sieh1997verify, breuer1997refinement}, 
users do not have to blindly trust a whole CPU but just the FPGA configuration modules and RTL verifiers.

\subsection{Hardware Modules and Root of Trust}

\begin{figure}[t]
	\begin{centering}
		\centering
		\includegraphics[width=0.48\textwidth]{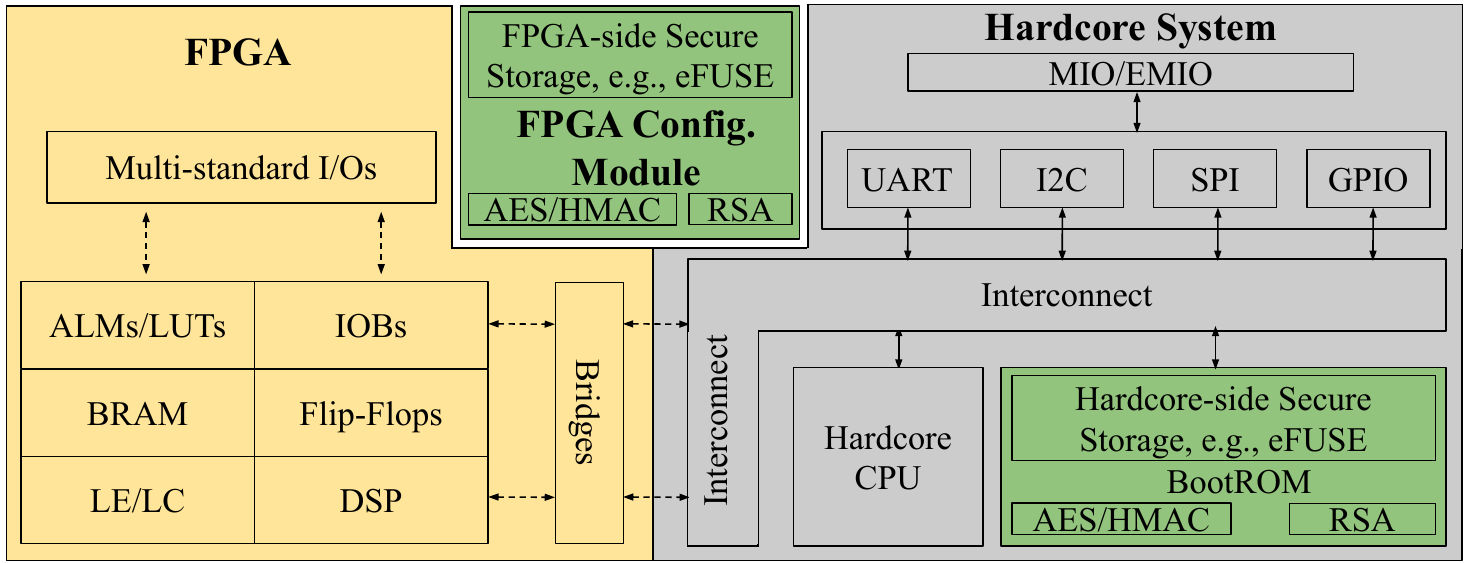}
		\vspace{-0.5cm}
		\caption{SoC FPGA architecture. The modules in green are the \emph{root of trust} for the hardcore system and FPGA, respectively.
		Note that \sysname only relies on the root of trust for FPGA. Solid lines represent hard-wired connections, whereas dashed lines represent configurable connections.}
		\label{fig:zynq_device}
	\end{centering}
	\vspace{-0.7cm}
\end{figure}

Figure~\ref{fig:zynq_device} shows the three main modules of a SoC FPGA.
A \emph{hardcore system} is formed around hard processors, such as the x86/64 processor on AMD EPYC and the Cortex-A processor on Xilinx Zynq-7000 SoC~\cite{ZynqTrm},
which includes a BootROM that contains the immutable BootROM code. 
In the \emph{secure boot} process of the hardcore system, the BootROM code serves as the \emph{Static Root of Trust for Measurement} (SRTM) and uses the keys stored in the hardcore-side secure storage, e.g., one-time programmable electronic fuse (eFUSE). 

An \emph{FPGA} can implement arbitrary systems, including softcore CPUs. 
To this end, the FPGA is composed of configurable logic blocks, Adaptive Logic Modules (ALMs), Lookup Tables (LUTs), flip-flops, etc.
In addition to the general fabric, the FPGA has Block RAMs (BRAM) to store data.
Note that BRAM is made of Static RAM (SRAM) on existing SoC FPGA platforms.
Compared to DRAM whose cells are made of capacitors and is vulnerable to cold-boot attacks due to the slow decay~\cite{halderman2009lest}, SRAM decays faster~\cite{rahmati2012tardis}.
The FPGA also includes Input/Output Blocks (IOB) for interfacing. 

An \emph{FPGA configuration module} configures the FPGA with a \emph{bitstream}. 
The module not only configures the hardware
but also loads optional software onto the BRAM.
Therefore, a bitstream can include: (i) hardware configurations, which are programmed in RTL hardware design description languages, such as Verilog and VHDL; and (ii) optional software, e.g., firmware, that runs on the hardware configurations. 
In the \emph{secure configuration} process of the FPGA, the FPGA configuration module (FCM) serves as the \emph{Root of Trust for Measurement} and uses the keys in the FPGA-side secure storage, e.g., BBRAM, eFUSE, to verify, decrypt, and configure an encrypted bitstream.
Note that the BootROM code cannot access the FPGA-side secure storage, and vice versa.

\subsection{Secure Configuration of FPGA}
A typical design and development flow of general-purpose computing platforms on SoC FPGA involves: 
(i) the development of the hardware on the FPGA, including designing the peripheral blocks and creating the connections. 
In this step, a developer can use and customize open-sourced and proprietary hardware IPs;
and (ii) the development of the software on the hardcore and softcore CPUs.

As aforementioned, the FPGA has a \emph{secure configuration} process, which is independent from the \emph{secure boot} process of the hardcore system. 
To facilitate the secure boot and secure configuration processes, cryptographic keys can be generated off the device and programmed to the corresponding secure storage on the device during manufacturing or through the JTAG interface if available.
These keys and the FPGA configuration module are the \emph{root of trust} of \sysname.
When a SoC FPGA device is powered on, the hardcore system boots. 
Privileged software, e.g., First Stage Boot Loader (FSBL) or OS, on the hardcore system can read a bitstream at any time from the persistent storage and send it to the FPGA configuration module, which verifies and configures the FPGA.

\section{System, Threat and Deployment Model}
\label{s:thread}

\textbf{System Model.} 
We assume the secure configuration process of FPGA discussed in \S\ref{s:back}. 
We assume the DRAM can be configured to connect to both of the hardcore system and FPGA, and peripherals can be connected to the FPGA without routing through the hardcore system.
The former enables the hardcore system and FPGA modules to communicate efficiently via shared memory, and the latter makes sure the software on the hardcore system cannot eavesdrop or tamper the data between the FPGA and peripherals.  
We assume the hardcore system and Direct Memory Access (DMA) masters cannot dump the content in the FPGA.
All of the assumptions are realistic in that they are the standard configurations on most commercial systems~\cite{parno2011bootstrapping}.
Even though recent research~\cite{ender2020unpatchable} discovered dumping content from the FPGA is possible due to some implementation bugs, it was not intended to be a feature.
We assume the cryptographic algorithms are secure.

\textbf{Threat Models.} We assume adversaries can compromise the hardcore system at boot-time or runtime, which means applications, kernel, and hypervisor are malicious.
The compromised software on the hardcore system can send arbitrary data to the firmware and SSAs in enclaves via shared DRAM regions and to the enclave hardware pins, such as interrupts. 
Adversaries can also perform cold-boot attacks to dump the content in DRAM.

For software running on the softcore CPUs, we first consider a baseline model (\modela) as the baseline design. 
We then consider \sysname under an enhanced attack model (\modelb). 
In \modela, the software in an enclave, including the firmware and SSA, are trusted and bug-free. 
The hardcore system cannot compromise the firmware or SSA at runtime, and remote attestation can be implemented in the firmware. 
This model is similar to the Arm TrustZone model where software-based attestation is trustworthy~\cite{ArmPSASM, ArmPSAattestation}. However, this model is not realistic as the firmware and SSAs may have bugs that can be exploited by REE inputs~\cite{cerdeira2020sok, machiry2017boomerang}. 
In \modelb, we assume that the firmware and SSAs are buggy and can be compromised. 
Therefore, measurement code and keys cannot be kept in the same address space as the firmware and SSAs. 
This model is similar to the Intel SGX and Arm CCA~\cite{li2022design} model where trusted hardware components of the CPU perform remote attestation. 
We do not consider the Time-Of-Check-Time-Of-Use (TOCTOU) attacks on hybrid remote attestation; however, we will discuss possible solutions in \S\ref{s:limit}.



\textbf{Key Management Model.}
\sysname provides the necessary mechanisms to build a secure system, which can integrate different deployment and key management models that are related to the user and application's policies and needs. 
In this paper, we discuss two key management models for local and cloud deployment scenarios. 
We assume that the FPGA configuration module securely stores a device key, such as AES ($k_{d}$) or RSA ($sk_{d}, pk_{d}$), which is used to encrypt/decrypt and sign/verify a bitstream. 
These device keys are unique to each device and are programmed in a secure storage using hardware interfaces with physical access, such as during manufacturing.
We also assume that developers can be identified by a developer key, such as AES ($k_{u}$) or RSA ($sk_{u}, pk_{u}$), which they use to encrypt and sign the SSAs. 
In \modela, the \firmware uses the developer key to decrypt and verify an SSA, and developer keys are embedded into the \firmware during the development stage. 
In \modelb, the developer keys are embedded in the trusted hardware-based module \hwloader for decryption and attestation.

\begin{figure*}[t]
	\begin{centering}
		\centering
		\includegraphics[width=.95\textwidth]{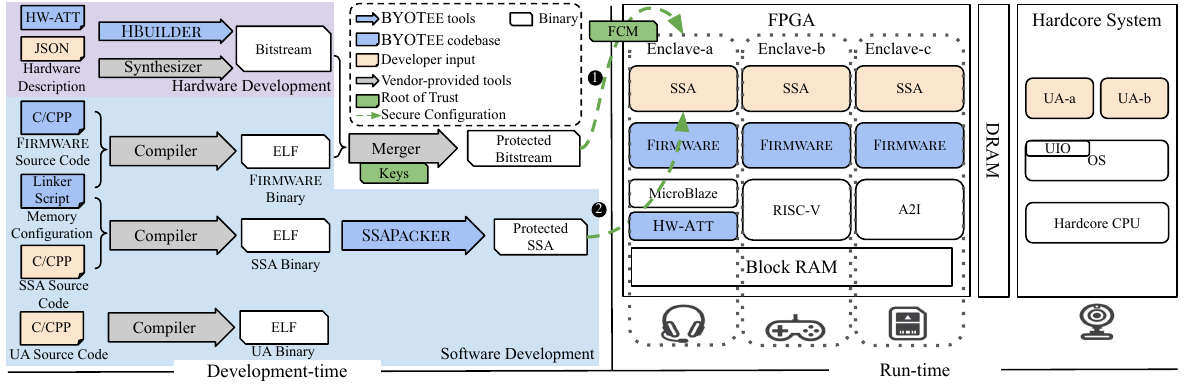}
		\vspace{-0.3cm}
		\caption{The architecture and workflow of the \sysname framework at development- and run-time. During development, \sysname and vendor-provided tools are used to generate the protected FPGA image and protected SSAs, which are loaded onto the FPGA (\Circled{\textbf{1}}) and enclaves (\Circled{\textbf{2}}), respectively. 
In this runtime architecture example, three enclaves with different hardware configurations, including softcore CPUs and peripherals, 
are presented. 
Untrusted Applications (UA) access the shared DRAM region through a userspace I/O interface (UIO).}
		\label{fig:BYOTee}
	\end{centering}
			\vspace{-0.4cm}
\end{figure*}

\section{\sysname Architecture}
\label{s:arch}
In this section, we first present the security and functional design goals of \sysname followed by an overview of its architecture and workflow.
We then illustrate hardware TCB customization, bootstrapping trust in enclaves, secure execution of SSA, and remote attestation mechanisms.

\subsection{Design Goals}
\label{subsec:designgoal}

\sysname provides isolated execution environments on-demand, which hardware debuggers and DMA-enabled devices cannot access.
With \sysname, users can use the exact and even formally verified RTL design and software needed for their applications.
\sysname has the following security and functional design goals: 

\goal{Customizable hardware TCB.} The hardware TCB of each enclave should be customizable, allowing for a minimum TCB that only includes the necessary hardware, e.g., peripherals, for the SSA and excludes other hardware on the system.
Please note that formally verifying the customized RTL hardware design~\cite{sieh1997verify, breuer1997refinement} is beyond the scope of \sysname.

\goal{Remote attestation mechanisms.} The mechanisms in \sysname should support protocols for remote verifiers to attest to the integrity of an enclave's hardware and software stacks. This includes the bitstream, \firmware, SSAs, and their inputs and outputs.

\goal{General-purpose execution environments.} \sysname should provide general-purpose execution environments, similar to SGX and TrustZone, and not limited to application-specific accelerators~\cite{hwang2021ambassy}. 
The SSAs can be implemented in any programming language, as long as they can be linked against the firmware.

\goal{Multiple isolated execution environments.} \sysname should provide multiple execution environments, similar to SGX, to ensure the confidentiality and integrity of the SSA running inside each environment. 
TrustZone and Ambassy~\cite{hwang2021ambassy} only provide a single execution environment, which can limit the flexibility of the system.

\goal{Circuit-level execution isolation.} \sysname should provide dedicated CPUs for each execution environment, ensuring that all hardware resources for each enclave are isolated from the REE and from other enclaves at the circuit level. This approach mitigates micro-architectural side-channel attacks, such as cache side-channel attacks, that are prevalent in CPU-sharing TEEs such as SGX and TrustZone.
However, please note that power side-channel attacks~\cite{zhao2018fpga} may still be possible, which will be discussed in~\S\ref{sec:analysis}.

\goal{Isolated path between SSA and peripherals.} An enclave should isolate the communication path between the SSA and peripherals from the hardcore system and other enclaves, preventing software-based eavesdropping and tampering.

\goal{Enclave-to-hardcore and Inter-enclave communication.} The SSA in an enclave should be able to communicate with software on the hardcore system and other enclaves. 
The inter-enclave communication should be isolated from the hardcore system and non-participating enclaves.

\goal{Allowing for minimum software TCB.} The firmware serving an SSA should only include necessary housekeeping libraries and drivers that are necessary for the SSA execution and exclude other software on the system.

\goal{Easy to use.} \sysname should be easy to use, especially for software developers who lack hardware programming experience. As a rule of thumb, developing an SSA should not take significantly more time and effort than developing a Linux application with the same functionality.

\subsection{\sysname Overview}
\label{subsec:overview}

Figure~\ref{fig:BYOTee} presents an overview of the architecture and workflow of the \sysname framework.
The \sysname tools and codebase mainly include the \hbuilder, \hwloader for the \modelb, \firmware, and \SSAPacker.
During the development stage, the \hbuilder generates synthesizer commands based on the SSA's needs specified in the developer's hardware description JSON input.
Then, the vendor-provided synthesizer, e.g., Xilinx Vivado~\cite{vivvado}, Intel Quartus Prime~\cite{quartus}, generates the bitstream file using the synthesizer commands.
The bitstream and \firmware binary are encrypted, signed, and packed by the vendor-provided merger, e.g., \verb|UpdateMEM| from Xilinx, into a protected bitstream.
The SSA binary is encrypted, signed, and packed by the \SSAPacker into a protected SSA.
When the bitstream is loaded onto the FPGA,
multiple enclaves can be created and \firmware starts running.
Then, an untrusted application can trigger the loading of a protected SSA into an enclave. 


\sysname meets \emph{G1}, \emph{G3}, \emph{G4}, \emph{G5}, and \emph{G6} by configuring the FPGA to build enclaves.
Enclaves are constructed with softcore CPUs, which provide a general-purpose computing environment (\emph{G3}).
Each enclave has its own set of hardware (\emph{G4}), including a softcore CPU (e.g., MicroBlaze, UltraSPARC), Block RAM, and peripherals.
Using FPGA routing, these hardware resources within an enclave are connected together, but isolated from the hard-core system and other enclaves (\emph{G5}).
The softcore CPU in an enclave is not time-shared with the hard-core system and other enclaves, mitigating cache side-channel attacks (\emph{G5}).
No additional hardware modules, such as debuggers, can be connected to an enclave unless explicitly specified by the developer (\emph{G1}).
Furthermore, all connections among these resources are isolated at the circuit level from the hard-core system and other enclaves, preventing eavesdropping and tampering (\emph{G6}).

Enclaves in \sysname use interrupts on softcore CPUs and shared physical memory regions on DRAM to communicate with the REE. 
In contrast, enclaves use interrupts and shared regions on the BRAM to communicate with each other. 
Since a shared BRAM region is only mapped in the address spaces of the participating enclaves, it is isolated from the REE and other enclaves, satisfying \emph{G7}.
The \sysname firmware can be customized and only consists of libraries, a HAL, and a loader for the SSA, meeting \emph{G8}.
To prove the integrity of enclaves, \sysname provides remote attestation mechanisms under two models, as required by \emph{G2}.
Additionally, \sysname provides an easy-to-use toolchain for developers to focus on SSA development, increasing the usability of \sysname and decreasing the chances of developer-induced misconfigurations, which satisfies \emph{G9}.

\subsection{Customizing Hardware TCB for Enclaves}
\label{subsec:customizeHWTCB}

To customize the hardware TCB, the developer designs enclaves using a hardware description language, e.g., Verilog or VHDL. 
The output is a hardware configuration bitstream file. 
To facilitate this step, \hbuilder takes developer-specified hardware description in JSON format as input (See an example in \S\ref{subsec:dev}), allocates hardware resources, and outputs a script, e.g., in Tcl format, that can be processed by a synthesis tool to generate the bitstream.
Each enclave's hardware description includes but not limited to: 
(i) a softcore CPU and its configurations, e.g., clock frequency, cache size, etc; 
(ii) the selection of software- or hardware-based attestation mechanism;
(iii) a corresponding debug IP to enable software debugging on the softcore CPU;
iv) its main BRAM memory address and size; 
(v) the address and size of the shared DRAM with the hardcore system; 
(vi) the address and size of the shared BRAM with other enclaves; 
and (vii) connected peripherals.
The \hbuilder assigns a contiguous address space of the BRAM to each enclave and connects the hardware components automatically.

\textbf{The \hwloader Module}.
If an enclave uses hardware-based attestation, \hbuilder will automatically connect the trusted hardware \hwloader module to it. The \hwloader is implemented in RTL and synthesized to a bitstream, and it can be formally verified. It is connected directly to the entire BRAM of the enclave and can operate on the enclave's BRAM directly. The \hwloader is responsible for decrypting an SSA, computing the measurements, and signing a measurement report. Cryptographic keys can be embedded in the \hwloader's own BRAM to prevent other hardware or software components from accessing them. 
In attestation under the \modelb, the \hwloader serves as the \emph{root of trust for measurement} and \emph{reporting}.


\subsection{Bootstrapping Trust in Enclaves}
\label{subsec:bootstrapTrust}

Software running on the hardcore system can configure the FPGA by sending a protected bitstream to the FPGA configuration module (FCM), which is a trusted hardware module.
The FCM verifies the bitstream using the device keys.
Upon a successful verification, the FCM decrypts the bitstream using the device key and configures the FPGA.
After this step, the softcore CPU, \hwloader in the \modelb, and the interconnections among hardware modules will be configured.
In addition, \firmware on the softcore CPU starts execution.
A measurement $m=H(bitstream)$ is generated by the FCM and placed on the shared DRAM region with the enclave for the future use of attestation report generation. 
In the \modela, the \firmware uses the developer keys, e.g., $k_{u}$ and/or $pk_{u}$, to decrypt and measure the SSA.
Allowed developer keys are embedded in the \firmware at the development stage.
Because the \firmware is encrypted at rest and only decrypted on the BRAM, the developer keys are secure.
Under the \modelb, \firmware can be compromised, so it requests the \hwloader to decrypt and verify the SSA.
Because the \hwloader is also encrypted at rest and only has the keys on its own BRAM at runtime, the keys are secure.

\subsection{Executing SSAs in Enclaves}
\label{subsec:execintee}

\begin{figure}[t]
	\begin{centering}
		\centering
		\includegraphics[width=0.47\textwidth]{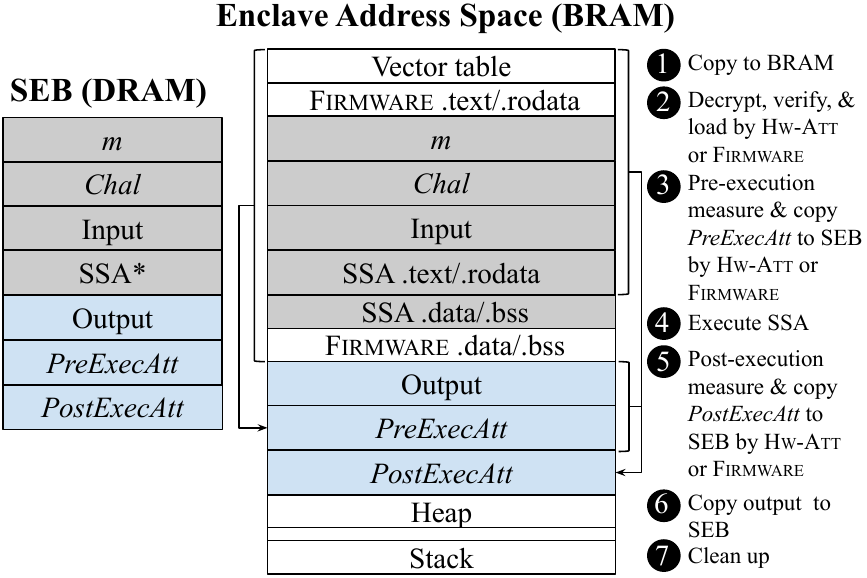}
		\caption{SSA Execution Block (SEB) layout, simplified enclave address space layout, and steps in executing an SSA.}
		\label{fig:sebctee}
	\end{centering}
\end{figure}

To create an SSA, the developer links the code against the \firmware. 
After launching an enclave, the \firmware initializes the softcore CPU and other components, then it waits for requests from the hardcore system.
Both the \firmware and SSA use a shared DRAM region in the SSA Execution Block (SEB) format as shown in Figure~\ref{fig:sebctee}, to have two-way data transmissions with UA on the hardcore system.
To initiate the transmission from the hardcore system to an enclave, UA on the hardcore system raises interrupts on the enclave's softcore CPU, which are handled by the \firmware.
\sysname defines three primitives through the \textsf{LdExec*} interrupts: (i) load and execute an SSA (\textsf{LdExec}); 
(ii) load and execute an SSA with pre-execution attestation (\textsf{LdExecPreAtt});
(iii) load and execute an SSA with post-execution attestation (\textsf{LdExecPostAtt}).
The softcore CPU interrupts can be implemented as GPIO interrupts in the enclave and memory-mapped to a DRAM address for the UA to access.
In this subsection, we focus on \textsf{LdExec}, and the other two primitives are discussed in \S\ref{subsec:attest}.

To execute an SSA on an enclave, the untrusted application first fills data into the SEB and raises a \textsf{LdExec*} interrupt.
As shown in Figure~\ref{fig:sebctee}, a SEB has regions for the encrypted and signed SSA (SSA*), input data for the SSA, output data from the SSA, a challenge $Chal$ from a remote verifier, a pre-execution ($PreExecAtt$), a post-execution attestation measurement ($PostExecAtt$) and other data.
When the \firmware receives a \textsf{LdExec*} interrupt, it copies SSA*, $Chal$, and input data in the SEB from DRAM to its own BRAM \Circled{\textbf{1}}.
The \firmware can disable \textsf{LdExec*} interrupts after data is copied.

Under the \modela,
the \firmware then decrypts and verifies the encrypted SSA* using the corresponding developer's keys \Circled{\textbf{2}}.
Under the \modelb,
the \firmware requests the \hwloader to decrypt and verify the encrypted SSA*. 
Upon the successful verification of the SSA's integrity, the \firmware loads sections of the decrypted SSA to the right locations and gives the control to the SSA \Circled{\textbf{5}}.
If there is an output, the SSA writes it in the output region on the BRAM, and yields the control of the softcore CPU back to the \firmware. 
The \firmware copies the output from the BRAM to the DRAM \Circled{\textbf{6}}. 
Finally, the \firmware cleans up all the input, output, and SSA-related regions on the BRAM and awaits new requests from the hardcore system \Circled{\textbf{7}}. 
While SSAs can execute concurrently on their own enclaves respectively, 
the \firmware also supports executing multiple SSAs sequentially or the same SSA multiple times on the same enclave without reconfiguring the FPGA but just re-initializing the enclave, e.g., flush the cache, 
clean up the BRAM \Circled{\textbf{7}}.

\subsection{Remote Attestation Mechanisms}
\label{subsec:attest} 

\sysname provides two attestation mechanisms, namely pre-execution and post-execution attestations. 
The former extends remote attestation of code and input integrity with bitstream, whereas the latter extends the output data integrity attestation~\cite{abera2019diat}.
Note that \sysname only provides the mechanism for attestation, which can support sophisticated attestation protocols.
With the help of trust bootstrapping discussed in \S\ref{subsec:bootstrapTrust}, the measurement mechanism not only captures the identity of the loaded SSA but also the bitstream, including hardware configurations and the \firmware. 
Note that it is critical to perform measurement on the BRAM since the DRAM can be changed asynchronously by the hardcore system.

\textbf{Software-based Attestation under the \modela}.
In pre-execution attestation, a verifier sends a cryptographic nonce as $Chal$, which is copied to the BRAM by the \firmware \Circled{\textbf{1}}.
After loading the SSA sections to the right addresses, the \firmware computes a measurement $PreExecAtt$ on the vector table, \firmware code and data, $m$, $Chal$, input data and SSA sections \Circled{\textbf{3}}, and copies the measurement to the DRAM.
Depending on scenarios and attestation protocol details, the \firmware can use a developer key or other shared keys to compute the measurement.
In post-execution attestation, 
after the SSA finishes execution \Circled{\textbf{4}} the \firmware computes a measurement $PostExecAtt$ on the vector table, \firmware code and read-only data, $m$, $Chal$, input data, output data generated by the SSA, SSA's read-only sections and $PreExecAtt$, and copies the measurement to the DRAM \Circled{\textbf{5}}.

\textbf{Hardware-based Attestation under the \modelb}.
The \firmware does not perform the measurements under the \modelb because it does not possess keys. 
Instead, it requests the \hwloader to compute the measurements. 
As discussed in \S\ref{subsec:customizeHWTCB} and shown in \S\ref{subsec:seceva}, the \hwloader is directly connected to the BRAM of the enclave, and it computes the measurements by reading the contents from the BRAM. 
The \hwloader then signs the measurements with the its keys and returns them to the \firmware or REE, which then sends the signed measurements to the remote verifier.

\subsection{Multiple Inputs to SSA}
\label{subsec:coninput}

In the case that the UA on the hardcore system needs to continuously send data to the SSA, e.g., not all input data is available at the beginning, the size of SEB is not big enough, etc., the UA writes the newly available input data in the input region inside the SEB, and it can use two mechanisms to notify the \firmware and SSA that new data is available. 
The first mechanism works for softcore CPUs that support priority interrupts. 
On such systems, \sysname defines a \textsf{NewData} interrupt, which UA can raise.
The \textsf{NewData} interrupt has a low priority so that it cannot interrupt the execution of the SSA.
Only after the SSA finishes execution and yields the control back to the \firmware, the \firmware can copy the input data from the DRAM to the BRAM, and gives the control to the SSA again.
On softcore CPUs without priority interrupts, the \firmware uses global variables to indicate whether new data is available in the input region to synchronize with the SSAs on the enclaves.

\subsection{Optional Multiple Protected SSA Sessions}
As an option, an enclave can also interleave the execution of multiple SSAs with proper hardware re-initialization, e.g., flush cache, reset all memory, etc.
To this end, \sysname defines two service primitives: (i) suspend and export the SSA state (\textsf{SusExp}); 
(ii) restore and execute a saved and encrypted SSA state (\textsf{ReExec}).
When a \textsf{SusExp} interrupt is raised, the \firmware copies the SSA context, e.g., general and system registers, onto the BRAM.
Then, the \firmware uses the developer key to encrypt and sign the saved SSA context and all of the SSA's writable memory regions, e.g., stack, .data, .bss, etc.
The encrypted blob is placed in the SEB output region for the UA to retrieve,
after which \firmware cleans up BRAM and awaits new requests. 
When a \textsf{ReExec} interrupt is raised, \firmware retrieves an encrypted blob from SEB. 
Upon a successful signature verification, \firmware loads the decrypted memory contents to the right locations, restores the registers, and resumes the SSA execution. 

\section{Applications and Developer's Perspective}
\label{s:dev}

In this section, we use four SSA examples to demonstrate how \sysname can secure real-world applications in several classes.
Then, we discuss how developers can easily develop and deploy enclave, SSAs, and UAs using the \sysname toolchain.

\subsection{\sysname Applications}
\label{sub:apps}

\textbf{Computational Applications.}
Computational applications take input from the hardcore system or other enclaves, perform the intended computational operations, and send the outputs back.
They represent computational tasks, such as encryption, decryption, machine learning-based classification, etc., that do not need peripherals.
\sysname protects such applications from code and data disclosure, memory corruption, and cache side-channel attacks from the hardcore system and other enclaves at runtime.   
We implemented an AES accelerator SSA (SSA-1) as an example for computational applications. 
To use SSA-1, a UA places the plaintext or ciphertext in the SEB and notifies the SSA. 
When the encryption or decryption is finished, SSA-1 places the outputs on the SEB.


\textbf{Peripheral-Interacting Applications.}
These applications interact with peripherals but do not communicate with other SSAs or the hardcore system. 
For instance, cyber-physical applications that read from sensors, make local decisions, and control an actuator fall under this category.
Besides the attacks \sysname protects the computational applications from, 
\sysname protects the paths between the SSA and peripherals from attacks. 
To demonstrate this protection, we developed an LED toggler SSA (SSA-2) that utilizes a button and an LED. 
Both the button and LED are solely connected to the enclave of SSA-2, rendering them inaccessible by the hardcore system or other enclaves.

\textbf{Peripheral- and Hardcore System-Interacting Applications.}
For demonstration, we developed a music player with digital rights management that guarantees the confidentiality, integrity, and authenticity of songs. 
This means (i) songs cannot be digitally disclosed, (ii) songs cannot be modified, 
and (iii) only songs that were protected can be played.

To this end, the music player system has three components:
(i) a trusted song protector (in Python with 160 SLOC), which is an offline component to encrypt and sign a song file (WAV format);
(ii) a UA (in C with 695 SLOC) running on the hardcore system, which provides a user interface to play, pause, resume, and stop a protected song.
The UA awaits for the user's commands, reads protected songs from storage, e.g., SD card, and sends them to the song playing SSA.
Because the protected song file is big (e.g., an original  77 seconds, 48KHz, and a single channel WAV file is around 33MB. 
The protected song file is several hundred bytes bigger.), the UA needs to continuously read the protected song file data from the storage and send it to the SSA;  
(iii) a song playing SSA (SSA-3) that authenticates, decrypts, and plays a song by sending the plaintext data of it to a hardware audio module.
The hardware audio module is only connected to the enclave running SSA-3.

We emphasize that any software solution that solely trusts SGX or TrustZone cannot meet the security requirements of this music player because
(i) there is no trusted I/O path between an SGX enclave and the hardware audio module; hence a malicious REE OS can breach the confidentiality, integrity, and authenticity of a song.
Some solutions, such as SGXIO~\cite{weiser2017sgxio}, attempt to address this issue but they add additional hardware, e.g., the hypervisor, into the TCB;
(ii) a TrustZone application must decrypt the song in DRAM before sending it to play; hence, vulnerable to cold-boot attacks.

\begin{figure}[!t]
	\begin{lstlisting}[label={l:CTEEHS},caption={An example hardware description defining three enclaves in JSON format.}] 
{"Enclaves": [
	{"Name": "Enclave-a",
	 "Processor": 
	 	{"Type": "MicroBlaze 32bit", "Debugging": "Enabled"},
	 "Memory Size": "512KB",
	 "Shared DRAM SEB": 
	 	{"Base": "0x20000000", "Size": "2MB"},
	 "Attestation": "Hardware"},
	{"Name": "Enclave-b",
	 "Processor": 
		{"Type": "VexRisc 32-bit",
	   	 "Data Cache": "16KB", "Instruction Cache": "16KB", 
	   	 "FPU": "F32", "Debugging": "Disabled"},
	 "Memory Size": "32MB",
	 "Shared DRAM SEB": {
	   "Base": "0x20000800", "Size": "128MB"},
	 "Attestation": "Software"}},
	{"Name": "Enclave-c",
	 "Processor": 
	 	{"Type": "A2I 64bit", "Data Cache": "64KB",
	   	 "Instruction Cache": "64KB",
	   	 "MMU": "Enabled", "MMU Page Size": "4KB",
	   	 "FPU": "AXU", "Debugging": "Disabled"},
	 "Memory Size": "64MB",
	 "Shared DRAM SEB": {
	   "Base": "0x20020800", "Size": "256MB"},
	 "Attestation": "Software"}}],
 "Peripherals": [
	{"Type": "AXI Gpio",
	 "Board Interface": "Btns 2bits",
	 "Access": ["Hardcore system", "Enclave-b"]},
	{"Type": "Uart Lite 8bit",
	 "Baud Rate": "115200",
	 "Access": ["Enclave-a"]},
	{"Type": "Dual Port BRAM Generator",
	 "Base Address": "0x1F0000", "Size": "2MB",
	 "Access": ["Enclave-a", "Enclave-c"]}]}
\end{lstlisting}
\end{figure}

\textbf{Distributed Applications.}
A distributed application consists of multiple inter-communicating SSAs running on different enclaves at the same time.
The SSAs communicate through a shared BRAM region.
\sysname not only protects each of the SSAs but also their communications from the hardcore system and other enclaves.
For demonstration, we developed an application that processes data in sequence with two SSAs.
SSA-1 first receives data from a UA, decrypts the data, outputs to the shared BRAM instead of DRAM, and the second SSA (SSA-4) takes the output of SSA-1 and performs a SHA512-HMAC signature verification.

\subsection{Developer's Perspective}
\label{subsec:dev}

\textbf{Creating Enclave Hardware.}
A developer can use the \hbuilder to design and create the hardware consisting of one or multiple enclaves.
Listing~\ref{l:CTEEHS} shows an example hardware description in JSON format of three enclaves. 
Enclave-a has a 32-bit MicroBlaze softcore CPU~\cite{mbsoft} and uses hardware-based attestation. 
Enclave-b has a 32-bit VexRisc softcore CPU~\cite{VexRiscv}, Enclave-c uses a 64-bit A2I softcore CPU~\cite{a2i}.
The softcore CPUs of Enclave-b and Enclave-c have FPUs, instruction, and data caches.
Hardware-based attestation and debugging is enabled on Enclave-a only, for which \hbuilder inserts the \hwloader and a debugging module. 
A DRAM region is reserved for the SEB of each enclave, respectively.
A UART peripheral is only connected to the Enclave-a and cannot be accessed by the hardcore system or the other enclave. 
Additionally, a GPIO peripheral is connected to both the hardcore system and Enclave-b but cannot be accessed by Enclave-a or Enclave-c.
Each enclave shares a DRAM region with the hardcore system for two-way enclave-to-hardcore System communication.
Enclave-a and Enclave-c can also use the shared BRAM region to communicate.

The developer uses \hbuilder to generate hardware configurations, which outputs scripts containing the synthesizer commands. 
The \verb|-d| parameter specifies the JSON configuration file, and \verb|-o| defines the output path.

\begin{lstlisting}[label={header}, language=bash, numbers=none] 
hardwareBuilder.py -d <CONFIG_JSON> -o <SCRIPT>
\end{lstlisting}

Then, the developer invokes the synthesizer tool with the script as input.
The \verb|-n| parameter specifies the name of the hardware project,
and the \verb|bf| parameter specifies the mode of operation, which includes generating bitstream, combining bitstream with \firmware, etc.
The output of the \hbuilder is a bitstream file specified by \verb|-o|.

\begin{lstlisting}[label={header}, language=bash, numbers=none] 
createFPGAImage -d <TCL> -n <PROJ_NAME> -bf <BUILD_FLAG> -o <FPGA_IMAGE>
\end{lstlisting}

\textbf{Creating Boot Images.}
After the \hbuilder, the developer uses the boot loader creation tool with the developer-defined boot image format, e.g., \verb|.bif|, and the protected FPGA image to create a deployable binary file.
\begin{lstlisting}[label={header}, language=bash, numbers=none] 
createBootImage <SYSTEM_BIF> <FPGA_IMAGE> -o <BYOTee_BIN>
\end{lstlisting}

\textbf{Creating SSAs and UAs.}
Software modules developed in any language that can be linked against the \firmware can be included in an SSA.
For example, SSAs developed in C can have their own \verb|main| functions with a declaration of \verb|int main()| \verb|__attribute__| \verb|((section (".text.ssa_entry")))|. 
Not all libc functions are available for the SSAs to use.
To move data among DRAM, BRAM, and peripheral memories, system-specific underlying mechanisms will be used. 
The \firmware provides a HAL with interfaces like \verb|BYOT_MemCpy| to replace the libc \verb|memcpy|. 
The UAs execute as unprivileged applications on the hardcore system 
and uses a UIO interface to communicate with the \firmware and SSA.
The developer uses the \SSAPacker to generate protected SSA binaries.

\begin{lstlisting}[label={header}, language=bash, numbers=none] 
SSAPACKER -d <SSA_BIN> -o <PROTECTED_SSA>
\end{lstlisting}

\section{Security Analysis}
\label{sec:analysis}

We conduct an informal security analysis of \sysname,
in which we discuss the attacks \sysname can and cannot defend. 


\textbf{Compromised Hardcore System.} 
Even if the hardcore system software is compromised at runtime, the attacker cannot access the data on/from enclave hardware resources, 
because they are in the isolated address space of the target enclave.
The attacker cannot breach the confidentiality of bitstream, SSA code and data at rest as well, because they are encrypted at build time.
The enclave-hardcore system's two-way communication is based on interrupts and the shared DRAM.
Malicious hardcore system software can raise the interrupt to the enclave to carry out a DoS attack.
Utilizing priority interrupts in sophisticated softcore processors, \sysname can prevent these attacks from the hardcore system side.

\textbf{Compromised \firmware and SSAs.} 
In the \modela, we assume \firmware and SSAs are bug-free and cannot be compromised.
Under the \modelb,
if \firmware and SSAs are compromised at runtime by malicious input sent by the hardcore system,
they can disclose information in the compromised enclave address space, including data on the BRAM and data from the connected peripherals.
But it cannot read data from the BRAM or peripherals of other enclaves. 
Therefore, the attack is confined within the compromised enclave.
The compromised \firmware and SSA cannot extract the keys from the \hwloader to forge measurement reports, since the BRAM of \hwloader is not connected to the enclave. 


\textbf{Malicious Hardware IPs and Peripherals.}
Malicious hardware IPs cannot be loaded since a bitstream is signed by a trusted developer and verified before loading.
Even if peripherals are malicious and send out rogue DMA requests to access sensitive memory regions, 
they are confined in the enclave they are assigned to.
Therefore, a malicious peripheral can only cause limited damages.

\textbf{Cold-boot Attack.}
While cold-boot attacks on DRAM at
room temperature are proven very effective~\cite{halderman2009lest}, attacks on SRAM without external power sources are less feasible~\cite{rahmati2012tardis}. 
Most data \sysname stores on the DRAM is either encrypted or does not need to be protected.
For instance, even if the SEB is located on the DRAM and subject to cold-boot attacks, 
the SSA*, which includes developer keys, is encrypted. 
Obviously, \emph{Chal}, \emph{PreExecAtt}, \emph{PostExecAtt} do not need to be protected.
It is, however, possible to dump the input and output fields of the SEB using cold-boot attacks on DRAM.
Other sensitive data, such as developer keys, plaintext SSA, the program states, are placed on an enclave or the \hwloader's BRAM.
Cold-boot attacks on BRAM are difficult because:
(i) the BRAM cells are hardware initialized during FPGA configuration in many SoC FPGA systems~\cite{bramInit};
(ii) even without initialization, the contents in BRAM decays faster~\cite{rahmati2012tardis}; 
(iii) BRAM is embedded on-chip and cannot be physically taken out, so attackers have to bypass software protections to dump its content.


\textbf{Cache Side-channel.} Because the CPU is time-shared between the REE and TEE in SGX and TrustZone, cache side-channel attacks are effective~\cite{brasser2017software, zhang2016return, cho2018prime, zhang2016truspy}. 
In \sysname, the REE on the hardcore system side and enclaves do not time-share any CPU resources; hence, there is no cache side-channel. 

\textbf{Power Side-channel.}
In FPGA-based remote power side-channel attacks, the attacker builds an on-chip ring oscillators-based power monitor to conduct power analysis on other modules on the same FPGA or a CPU on the same SoC~\cite{zhao2018fpga}.
\sysname cannot mitigate these attacks but can prevent them by only loading authenticated and trusted enclave bitstreams that do not have a power monitor.

\textbf{Other Side-channels.}
When multiple enclaves reside on the same SoC FPGA, they share FPGA hardware resources.
Therefore, it is possible to conduct other sharing-based side-channel attacks, such as FPGA long wire-based attacks~\cite{giechaskiel2018leaky, ramesh2018fpga}.
Similar to power side-channel attacks, \sysname cannot prevent these attacks directly.

\section{Implementation and Evaluation}
\label{s:impleval}

We present an implementation of the \sysname framework for the Xilinx SoC FPGA and evaluate it on a low-end Digilent Cora Z7-07S development board ($\approx$\$130). 

\begin{figure*}[!t]
	\centering
	\subfloat[Enclave-1 w/ debugger]{\includegraphics[width = 0.19\linewidth]{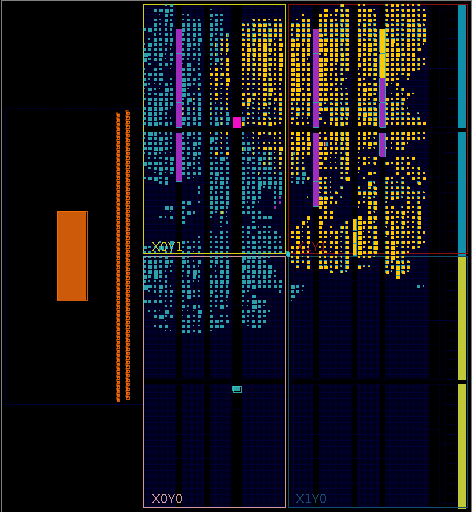}}\hfill
	\subfloat[Enclave-1 wo/ debugger]{\includegraphics[width = 0.19\linewidth]{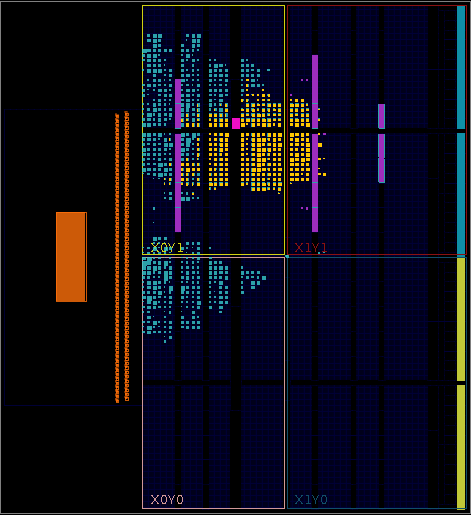}}\hfill
	\subfloat[Enclave-2 wo/ debugger]{\includegraphics[width = 0.19\linewidth]{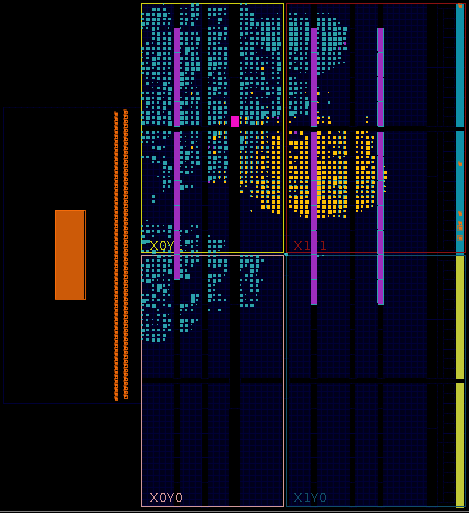}}\hfill
	\subfloat[Enclave-3 wo/ debugger]{\includegraphics[width = 0.19\linewidth]{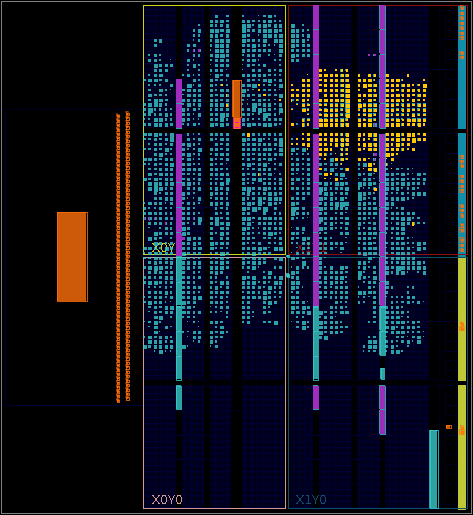}}\hfill
	\subfloat[Enclave-1 and Enclave-4 wo/ debugger]{\includegraphics[width = 0.19\linewidth]{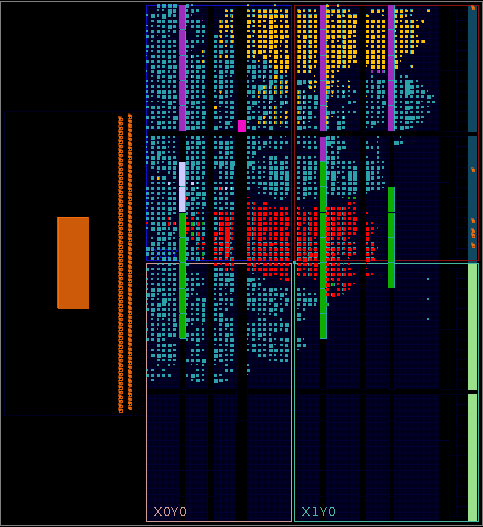}}    
	\vspace{-0.3cm}    
	\caption{Resource footprints of the enclaves (software-based attestation profile) with MicroBlaze softcore CPUs for the example SSAs on the Cora Z7-07S.
		The yellow and red portions represent CPU cells.
		The purple portion represents BRAM cells.
		The pink rectangle represents I/O ports. In (d), the rectangle on top of the I/O ports represents an analog to digital conversion module. The blue portions represent all other IPs, such as debugging modules, interconnects.}
	\label{fig:corr-res-iml}
	\vspace{-0.3cm}
\end{figure*}

\subsection{Experiment Environment}

\begin{table*}[t]
	\centering
	\setlength\tabcolsep{.75ex}
	\small
	\caption{Resource utilization of enclaves (software-based attestation profile) for the example SSAs on Cora Z7-07S}
	\vspace{-0.2cm}
	\label{t:resApps}
	\begin{tabular}{c|c|c|c|c|c|c|c|c}
		\hline
		 & \multicolumn{2}{c|}{Enclave-1}   & \multicolumn{2}{c|}{Enclave-2}  & \multicolumn{2}{c|}{Enclave-3}& \multicolumn{2}{c}{Enclave-4}  \\ \hline
		Resource & w/ debugger   & w/o debugger  & w/ debugger   & w/o debugger & w/ debugger   & w/o debugger & w/ debugger   & w/o debugger \\ \hline
		LUT     & 5,255 (36.5\%)      &   2,232 (15.5\%)    & 6,385 (44.3\%) &  3,291 (22.9\%) & 10,781 (74.9\%) & 6,778 (47.1\%) & 5,302 (36.8\%) & 3,130 (21.7\%) \\ \hline
		LUTRAM   &  419 (7.0\%)        &    211 (3.5\%)    &   507 (8.5\%) & 282 (4.7\%)  & 725 (12.1\%) & 427 (7.1\%) & 319 (5.3\%) & 145 (2.4\%)\\ \hline
		Flip-flop    & 5,245 (18.2\%)       &  2,259 (7.8\%)    &     6,759 (23.5\%)  & 37,64 (13.1\%) & 11,363 (39.5\%)&   7,721 (26.8\%) & 5,497 (19.1\%) & 3,014 (10.5\%)\\ \hline
		BRAM  & 18 (36.0\%)  & 16 (32.0\%)    &     34 (68.0\%) & 32 (64.0\%)   &  48 (95.0\%)   & 45.50 (91.0\%)   & 28 (56.0\%)  & 26 (52.0\%)\\ \hline
		DSP    & 3 (4.5\%) & 0 (0.0\%)&  3 (4.5\%)     &   0 (0.0\%) & 3 (4.5\%)  & 0 (0.0\%) & 3 (4.5\%) & 0 (0.0\%) \\ \hline
		IOB     & 0 (0.0\%) & 0 (0.0\%)   &  6 (6.0\%)   & 6 (6.0\%)   &   28 (28.0\%) & 28 (28.0\%) & 0 (0.0\%) & 0 (0.0\%) \\ \hline
	\end{tabular}
	\vspace{-0.25cm}
\end{table*}

The Cora Z7-07S board has a single-core 667MHz Arm Cortex-A9 processor with 512MB DDR3 memory, 32KB L1 cache, 512KB L2 cache and a Xilinx Zynq-7000 FPGA.
The Zynq-7000 FPGA has 3,600 logic cells, 14,400 LUTs, 6,000 LUTRAM, 28,800 flip-flops, a 225KB BRAM, 66 Digital Signal Processing (DSP) slices, and 100 IOBs.
The development board also has an SPI header, two push-buttons, two RGB LEDs, a microSD card slot, and two Pmod connectors.
We connect a Pmod I2S2 audio input and output device~\cite{pmod} to the board for SSA-3 evaluation.
Figure~\ref{fig:z7} shows the top and bottom view of the board with the connected audio device.

\subsection{\sysname Implementation}
We implemented the \sysname infrastructure and toolchain, which include \hbuilder, \hwloader for the \modelb, \SSAPacker, and \firmware, for the Xilinx SoC FPGA.
The \hbuilder was developed in Python (2.5K SLOC).
\hwloader can be developed in VHDL or C on a softcore CPU. 
In our implementation, \hwloader includes 1400 SLOC C code.
The \SSAPacker includes Python (63 SLOC) and C code (420 SLOC).
The \firmware was developed in C and has an SSA loader and cleaner (1.1K SLOC), an attestation module for the \modela (333 SLOC), an interrupt initialization and handling module (101 SLOC), and a linker script (212 lines).
The \firmware is linked against the vendor-provided HAL (7.9K SLOC) and libraries, e.g., libc (1.2MB), etc.
The \firmware, especially the HAL, can be customized to reflect an SSA's needs.
Our implementation uses AES-256 for SSA encryption and SHA512-HMAC to protect the integrity and authenticity of SSAs.
We use the BLAKE2~\cite{Blk2} hash algorithm to implement the pre-execution- and post-execution-attestations.
On the hardcore system side, a userspace I/O interface is used for the UAs to access the shared DRAM regions between the hardcore system and FPGA.
The \sysname toolchain also includes scripts to automate the steps from synthesizing the hardware, compiling SSAs and \firmware, and formatting the SD card with partitions. 

\begin{table}[t]
	\centering
	\setlength\tabcolsep{1.4ex}	
	\small
	\caption{Size of the example SSAs' software TCB}
	\label{t:FwLocRes}
	\scalebox{.95}{	
	\begin{tabular}{c|c|c|c|c|c|c|c}
		\hline   
		 & \multicolumn{2}{c|}{SSA} & \multicolumn{5}{c}{Corresponding \firmware} \\
			\hline   
		& SLOC & Bytes & SLOC & .text & .data & .bss & Total \\
		\hline 
		SSA-1 & 717 & 12,892& 3,143 &   27,296    &  3,236    &  448    &      30,532 \\
			\hline   
		SSA-2 & 346 & 2,868 & 3,532&  30,748     &  2,800  &   440   &  33,988   \\
		\hline   
		SSA-3 & 1,029 & 20,380 & 9,698 &     57,142  &    4,308  &    635  &  62,085   \\
		\hline 
		SSA-4 & 622 & 31,088 & 3,235 & 28,377   &   3,608   &   528  &  35,748  \\
		\hline 
	\end{tabular}
	}
    \vspace{-0.3cm}	
\end{table}

\begin{table*}[t]
	\centering
	\setlength\tabcolsep{1.15ex}	
	\small
\caption{Performance Evaluation of \firmware on MicroBlaze CPU (Software-based attestation; Time in Milliseconds; Size in Bytes).
The experiments demonstrate that SSA-3 efficiently verify, decrypt, and play 48KHz WAV on a low-end softcore CPU.}
	\label{t:all}
	\begin{tabular}{c|c|c|c||c|c|c|c|c|c||c|c}
		\hline   
		&  Binary & Input & Output & Loading & Decryption & Integrity and & pre-execution & post-execution  & Cleaning & Suspend & Restore \\   
		&  size & size & size &  & & authenticity & attestation & attestation & up & and & and\\
		&  &  & &  & & verification & & & & export & execute\\
		 \hline  		
		SSA-1 & 12,892 & 64 & 64 & 1.39 & 2784.56 & 118.54 & 153.94  & 154.79 & 0.54 & 3694.55 & 3609.65\\ \hline  
		SSA-2 &  2,596 & N/A & N/A & 0.30 & 579.11 & 29.15 & 30.90 & 30.90 & 0.11 & 741.71
		 & 729.69 \\ \hline  
		SSA-3 & 20,152 & 128
		& 132 & 2.17 & 4414.32 & 185.30 & 258.30 & 260.50 & 0.90 & 5787.97 & 5638.76 \\ \hline 
	\end{tabular}
\end{table*}

\begin{table}[t]
	\centering
	\small
	\caption{Performance evaluation of Embench-IoT on softcore MicroBlaze CPU (Version 10.0, 100MHz, no Cache, no FPU) and hardcore Cortex-M4 CPU (16MHz, no Cache, no FPU, officially reported performance from~\cite{embencm4}) in Milliseconds}
	\label{t:embench}
	\begin{tabular}{l|p{3cm}|c|c}
	\hline   
	 Application & Description & M4~\cite{embencm4} & MicroBlaze \\ \hline 
	 aha-mont64 & Modulo generator& 4,004 &  501 \\ \hline
	 crc32 & 32 bit error detector & 4,010 & 193\\ \hline 
	 huffbench & Data compressor & 4,120 &  111\\ \hline 
	 minver&Floating point matrix inversion&3,998 & 327\\ \hline
	 nettle-aes & Low level AES library & 4,026 & 245\\ \hline
	 nsichneu & Computes permutation&4,001 & 237\\ \hline
	 primecount &Prime counter & n/a & 193\\ \hline
	 sglib-combined& Sort, search, and query on array, list, and tree &3,981 &  189\\ \hline
	 slre&Regex matching&4,010&113\\ \hline
	 statemate&Car window lift control & 4,001 & 139\\ \hline
	 tarfind&Archive file finder& n/a &163
	  \\ \hline
	 ud & Matrix factorization & 3,999 & 343\\ \hline 
	 \multicolumn{2}{c|}{Geometric Mean}&4,015&208\\ \hline
	\end{tabular}
\end{table}

\subsection{Customized Enclaves for the Example SSAs}
We specified the hardware description for the SSAs in~\S\ref{sub:apps} and used the \hbuilder and synthesizer to generate the bitstream. 
All the enclaves are configured with a 32-bit Microblaze CPU (version 10.0, 100MHz, no instruction/data cache, no FPU). 
The Enclave-1, Enclave-2, Enclave-3 have a 128KB BRAM, whereas Enclave-4 has a 32KB BRAM as their main memory.
The peripherals that belong to an enclave are connected through a dedicated AXI Interconnect IP.
Figure~\ref{fig:corr-res-iml} shows the footprints of the four hardware designs on the Z7-07S device.
These figures demonstrate the configurable nature of the \sysname hardware TCB and the isolation at circuit level of the enclaves from each other and from the hardcore processor. 
Figure~\ref{fig:corr-res-iml}(a) shows the hardware design with a debugger, 
whereas all other designs do not have a debugger for the minimum hardware TCB.
Table~\ref{t:resApps} presents each enclave's hardware TCB and resource utilization on the Cora Z7-07S board with and without a debugger IP. 
As the table shows, the debugger IP significantly increases the resource utilization of an enclave as it uses three DSP slices, two BRAMs, etc.
Since SSA-1 and SSA-4 do not use peripherals, Enclave-1 and Enclave-4
do not have any IOB.
Figure~\ref{fig:corr-res} (in Appendix) shows the block diagrams of the enclaves generated by the \hbuilder for the four example SSAs.

%
%

\subsection{Security Evaluation}
\label{subsec:seceva}
\begin{figure}[!t]	
	\begin{centering}
		\centering
		\includegraphics[width=0.48\textwidth]{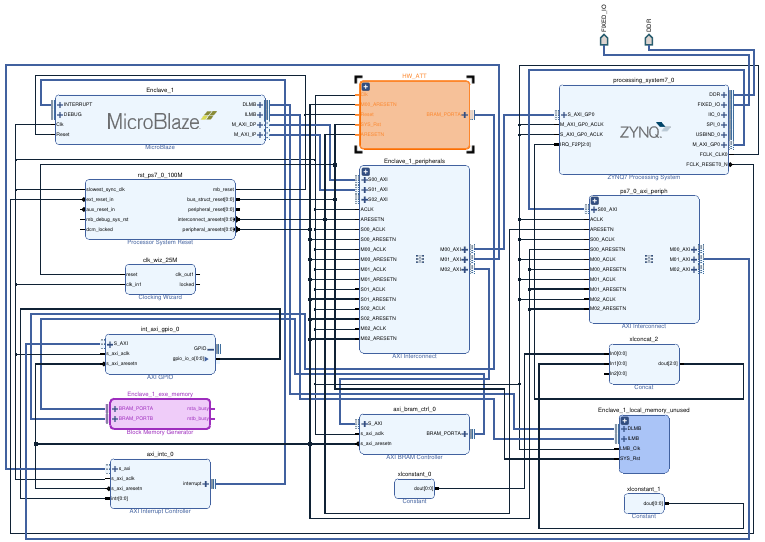}
		\cprotect\caption{Block diagram of SSA-1 with the \hwloader module (hardware-based attestation profile).}
		\label{fig:hw-att}
		\vspace{-.4cm}	
	\end{centering}
\end{figure}


\textbf{Circuit-level Execution Isolation}.
As shown in Figure~\ref{fig:corr-res}(e), the two enclaves for SSA-4 are isolated at the circuit level from each other and the hardcore system (\verb|processing_system7_0|).
The figure shows Enclave-1's instruction memory controller (\verb|Enclave_1| \verb|_ILMB|), data memory controller (\verb|Enclave_1_DLMB|), and memory generator (\verb|blk_mem_gen_0|) are only connected to the \verb|Enclave_1| CPU,
where \verb|Enclave_4_local_memory| (the combination of two memory controllers and one generator) is only connected to the \verb|Enclave_4| CPU.
To share a BRAM region for communication, each enclave has another memory controller, i.e., \verb|share_axi_bram| \verb|_ctrl_0| and \verb|share_axi_bram|\verb|_ctrl_0|, which is connected to the shared memory generator (\verb|share_blk_mem_gen_1|).



\textbf{Isolated Path to Peripherals}.
The hardware design in \sysname ensures isolated paths between SSAs and peripherals. 
As shown in Figure~\ref{fig:corr-res}(d), the I2S output audio peripheral (\verb|i2s_output_1|) for SSA-3 is only connected to its enclave but is isolated at the circuit level from the hardcore system.

\textbf{Hardware-based Attestation}.
Figure~\ref{fig:hw-att} shows the block diagram of the SSA-1 with hardware-based attestation.
Compared with Figure~\ref{fig:corr-res}(b), which uses software-based attestation, 
we can see the \hwloader is connected to the BRAM of Enclave-1 (\verb|Enclave_1_exe| \verb|_memory|).


\textbf{Software TCB Size}.
Table~\ref{t:FwLocRes} presents the size of the software TCB for the four example SSAs and their corresponding \firmware.  
As the table shows, the size of \firmware increases as the SSA gets more complicated and needs more services.
Nevertheless, the runtime software TCB (SSA and \firmware combined) of SSA-3, which is a functional digital right management music player, has only 10,727 SLOC, representing a significant software TCB reduction from its counterpart implemented as a TrustZone, e.g., TF-M~\cite{TFMsrc} has over 117K SLOC, or SGX application, e.g., the Gramine library OS~\cite{gramine} has 83K SLOC. 

\begin{figure}[!t]
	\centering
	\subfloat[0s 20$^{\circ}$C]{\includegraphics[width = 0.19\linewidth]{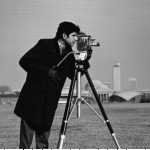}}\hfill
	\subfloat[5s 20$^{\circ}$C]{\includegraphics[width = 0.19\linewidth]{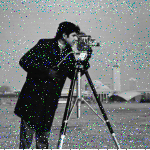}}\hfill
	\subfloat[15s 20$^{\circ}$C]{\includegraphics[width = 0.19\linewidth]{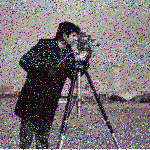}}\hfill
	\subfloat[20s 20$^{\circ}$C]{\includegraphics[width = 0.19\linewidth]{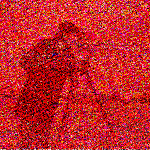}}\hfill
	\subfloat[30s 20$^{\circ}$C]{\includegraphics[width = 0.19\linewidth]{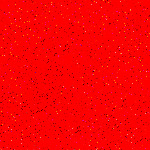}}  \\
	\subfloat[0s -18$^{\circ}$C]{\includegraphics[width = 0.19\linewidth]{images/0.png}}\hfill
	\subfloat[1m -18$^{\circ}$C]{\includegraphics[width = 0.19\linewidth]{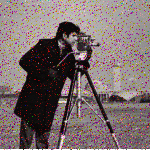}}\hfill
	\subfloat[10m -18$^{\circ}$C]{\includegraphics[width = 0.19\linewidth]{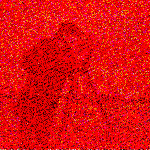}}\hfill
	\subfloat[13m -18$^{\circ}$C]{\includegraphics[width = 0.19\linewidth]{images/20.jpg}}    \hfill
	\subfloat[BRAM initialization]{\includegraphics[width = 0.19\linewidth]{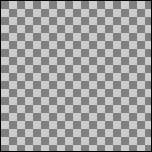}}\hfill
	\vspace{-0.3cm}
	\caption{Visualizing cold-boot attacks on DRAM and BRAM on the same Z7-07S board. 
	We loaded a bitmap image (150$\times$150 pixel; 90.1kB) on the DRAM and BRAM. 
	The reconstructed image from the fully decayed DRAM is red because half of the cells are 1s and the other half are 0s. The image from the BRAM is transparent because it is initialized to 0s.}
	\label{fig:cold-boot}
	\vspace{-.3cm}
\end{figure}

\textbf{Cold-boot Attacks on DRAM and BRAM}.
Cold-boot attacks on DRAM are a serious problem, especially when an attacker has physical access to the device.
We evaluated the feasibility of cold-boot attacks on DRAM and BRAM on the same board.
In these experiments, we loaded a bitmap image (150$\times$150 pixel; 90.1kB) and measured the DRAM decay at room temperature (20$^{\circ}$C/68$^{\circ}$F) and -18$^{\circ}$C/0$^{\circ}$F after power reset (0 second) and losing power for different intervals, e.g., 30 seconds, 13 minutes. 
We dumped the content of BRAM, for which the Xilinx Zynq-7000 FPGA has a non-bypassable hardware initialization mechanism after power up to clear all the bits to 0s.
As we discussed in~\S\ref{sec:analysis}, even if the BRAM is not initialized, cold-boot attacks on it are more difficult than on DRAM. 
Figure~\ref{fig:cold-boot} visualizes the cold-boot attack results, which confirms cold-boot attacks on DRAM are feasible but not on BRAM.

\subsection{Performance and Power Evaluation}

We evaluate the performance of the low-end MicroBlaze-powered enclaves, which provides a lower-bound performance estimation of available softcore CPUs.  
We evaluate the performance using 12 Embench-IoT benchmark applications~\cite{embenchIOT} and
evaluate the \sysname software performance by measuring the time cost of different \firmware operations.

\textbf{Benchmark Performance Evaluation}.
To show the performance of the MicroBlaze softcore compared to the Cortex-M4 hardcore, we use 12 applications from the Embench-IoT benchmarks~\cite{embenchIOT}.
As Table~\ref{t:embench} shows, the applications run comparatively faster on the low-end MicroBlaze softcore CPU than the hardcore Cortex-M4.
For better performance, users can choose more advanced softcore CPUs. 

\textbf{\firmware Performance}.
We evaluate the time \firmware spends on the loading, decrypting, integrity and authenticity verification, attestation, cleaning up, suspending, and restoring operations of three SSAs.
As Table~\ref{t:all} shows,
the time spent by \firmware is linear to the size of the SSA and its data.
To copy the protected SSA and its input from DRAM to BRAM, \firmware running on the Z7-07S spends around 1.07 ms for every 10,000 bytes.
To decrypt the protected SSA and its data using 256-bit CBC mode AES, \firmware spends around 2182 ms for every 10,000 bytes.
The integrity and authenticity verification costs around 93.43 ms for every 10,000 bytes.
The BLAKE-based pre-execution and post-execution attestations cost around 124.77 ms for every 10,000 bytes.
Cleaning up BRAM takes around 0.45 ms for every 10,000 bytes.
The SHA512-HMAC and AES 256-bit with CBC mode based suspending and restoring cost roughly 2834 ms for every 10,000 bytes.

\textbf{Power Consumption.}
The Xilinx Vivado (2017.4) tool provides the estimated power consumption of the hardcore CPU and example enclaves.
The 667MHz Cortex-A9 hardcore CPU uses around 1,255 mW. 
The SSA-1 enclave, which includes a 100MHz MicroBlaze, BRAM, etc., uses 66 mW. 
The SSA-2 consumes 72 mW, SSA-3 consumes 205 mW, and SSA-4 consumes 117 mW.
The hardware-based attestation profile of SSA-1 consumes 101 mW. 
For reference, a 16MHz Cortex-M4 consumes 0.66 mW~\cite{CrtxM4}.

\section{Related Work}
\label{s:rel}

\begin{table}[!t]
	\center
	\setlength\tabcolsep{1.5ex}	
	\small
	\scalebox{1}{	
		\begin{threeparttable}
			\begin{tabular}{l|p{3.2ex}|p{0.01ex}p{0.01ex}p{0.01ex}p{0.01ex}p{0.01ex}p{0.01ex}p{0.01ex}p{0.01ex}p{0.01ex}p{0.01ex}p{0.01ex}p{0.01ex}} \hline
				\multicolumn{1}{c}{Projects} & \multicolumn{1}{c}{} & \multicolumn{12}{c}{Benefits} \\ \hline 
				& \rotatebox{90}{Underlying Hardware Primitive}   & \rotatebox{90}{Customizable CPU and Memory (\emph{G1})} & \rotatebox{90}{Customizable Peripheral Connections (\emph{G1})}& \rotatebox{90}{Remote Hardware Attestation (\emph{G2})} & \rotatebox{90}{Remote Software Attestation (\emph{G2})} & \rotatebox{90}{Post-Execution Attestation (\emph{G2})}  & \rotatebox{90}{Multiple TEEs (\emph{G4})}   & \rotatebox{90}{Cache Side-channel Attack Resistant (\emph{G5})} & \rotatebox{90}{Concurrent TEE and REE Execution (\emph{G5})} & \rotatebox{90}{TEE with Dedicated Hardware (\emph{G5, G6})} & \rotatebox{90}{Allowing for Minimum Software TCB (\emph{G8})} & \rotatebox{90}{Cold-boot Attack Resistant} & \rotatebox{90}{Deployable on Commodity Devices} \\ \hline 
				Flickr~\cite{mccune2008flicker} & S/T  & & & &\cmark & \cmark & &  & &   &\cmark & & \cmark \\		
				TrustVisor~\cite{mccune2010trustvisor} & S/T  & & && \cmark & \cmark & &   &&  &\cmark & & \cmark \\	
				Haven~\cite{haven} & X & & & &\cmark & \cmark & \cmark&   & &   & & & \cmark \\	
				SGXIO~\cite{weiser2017sgxio} & X+H  & & \cmark& &  &  & \cmark &  & &  &  &  & \cmark \\
				SGX-FPGA~\cite{xia2021sgx} & X+F  & & \cmark&\cmark &\cmark &  & \cmark & & &  &  &  & \cmark \\	
				KeyStone~\cite{Keystone2020} & R  & & & & \cmark &  & \cmark &  & &  &  &  & \cmark \\			
				Sanctum~\cite{costan2016sanctum} & R & & \cmark & &  \cmark &\cmark & \cmark &  \cmark  & &   &  & & \\		
				{\scshape Cure}~\cite{bahmani2021cure} & R & & \cmark & &   & & \cmark &\cmark   &  &  &  & & \cmark \\	
				Composite Encl.~\cite{schneider2020pie} & R & & \cmark &\cmark & \cmark  & & \cmark &   &  & &  & & \cmark \\			
				{\scshape Sanctuary}~\cite{brasser2019sanctuary} & A & &  & & \cmark  & & \cmark &  &  &  & \cmark & & \cmark \\	
				TrustICE~\cite{sun2015trustice} & A & &  & &\cmark  & \cmark & \cmark &    &  & &  & & \cmark \\
				vTZ~\cite{hua2017vtz} & A+H  & &  &  & &  & \cmark &    & &  &  & & \cmark \\	
				Ambassy~\cite{hwang2021ambassy} & A+F  & $-$ & \cmark &\cmark &\cmark  & && \cmark & \cmark & $-$& $-$  & \cmark & \cmark \\		
				uTango~\cite{oliveira2021utango} & M  & & & &\cmark & &\cmark&   &&  & & & \cmark \\
				Graviton~\cite{volos2018graviton} & G & & & &\cmark& &\cmark&  & \cmark & & &\cmark & \\
				StrongBox~\cite{deng2022strongbox} & G & & & &\cmark& &\cmark&  & \cmark & & &\cmark & \cmark\\
				
				HECTOR-V~\cite{nasahl2020hector} & N  &  &\cmark & & & & \cmark & \cmark& \cmark& \cmark & \cmark& \cmark &  \\ 
				TEEOD~\cite{teeod} & F  & \cmark & & & & & \cmark & \cmark& \cmark& \cmark & \cmark&  & \cmark \\ 
				Sancus~\cite{noorman2013sancus} & N  &  & &&\cmark  & & & & &  &\cmark &  & \\ 
				SecureBlue++~\cite{boivie2012secureblue++} & -  &  & & & & & & &  & & \cmark&  & \cmark\\ 
				TrustLite~\cite{koeberl2014trustlite} & N  &  & & & \cmark &  & \cmark& &  & &\cmark &  &\\ 
				SMART~\cite{eldefrawy2012smart} & N  &  & &  &\cmark & \cmark & & &  & & &  &\\ 
				MyTEE~\cite{hanmytee}& H  &  & \cmark &  & && & &  &&\cmark &  &\cmark\\ 
				MeetGo~\cite{oh2021meetgo}& F  & $-$ &  & \cmark & $-$& &\cmark &$-$& \cmark &$-$& $-$& \cmark &\cmark\\ 
				\hline
				\sysname & F &\cmark &\cmark & \cmark & \cmark & \cmark & \cmark & \cmark & \cmark & \cmark & \cmark & \cmark & \cmark \\ 
				\hline                                                                   
			\end{tabular}
			\begin{tablenotes}
				\item[] S: AMD Secure Virtual Machine extension, T: Intel Trusted eXecution Technology, H: Hypervisor, A: Arm Cortex-A TrustZone, M: Arm Cortex-M TrustZone, X: Intel SGX, F: FPGA, R: RISC-V, G: GPU, N: New hardware design. $-$: not applicable.
			\end{tablenotes}  
		\end{threeparttable}
	}	
	\caption{Comparing the security goals and benefits of \sysname with other software- and hardware-based TEE solutions}
	\label{tab:comparison}
	\vspace{-0.4cm}	
\end{table}

Many software- or hardware-based solutions have been proposed to address one or more limitations of existing TEEs. 
Among them, TEEOD~\cite{teeod} is most related.
Compared to TEEOD, 
\sysname offers additional security features, such as trust bootstrap, software- and hardware-based attestation.
Table~\ref{tab:comparison} highlights the advantages of \sysname and compares it to related work. 
Moreover, we discuss previous efforts on addressing the single TEE issue, isolated I/O paths, and the limitations of other hardware-based solutions.

\textbf{The Single TEE Issue of TrustZone.}
vTZ~\cite{hua2017vtz} provides each virtual machine with a virtualized TEE by running a monitor within the secure world. 
{\scshape Sanctuary}~\cite{brasser2019sanctuary} 
utilizes the memory access controller to provide multi-domain isolation.
TrustICE~\cite{sun2015trustice} creates multiple
computing environments in the normal domain and runs a monitor in the secure world. 
uTango~\cite{oliveira2021utango} use the secure attribution unit of Cortex-M to create multiple secure execution environments.
On RISC-V, KeyStone~\cite{Keystone2020} utilizes the Physical Memory Protection (PMP) feature to create multiple enclaves.
The TEE and REE in these solutions time-share the CPU and other hardware resources, resulting in side-channel attacks.

\textbf{Isolated I/O Paths and Mitigating Side-channel Attacks.}
{\scshape Cure}~\cite{bahmani2021cure} enables the exclusive assignment of system resources to single enclaves.
Composite Enclaves~\cite{schneider2020pie} builds on top of KeyStone~\cite{Keystone2020} and extends the TEE to several hardware components.
HECTOR-V~\cite{nasahl2020hector} uses a dedicated processor as a TEE with configurable peripheral permissions. 
{\scshape Cure}, Composite Enclaves, and HECTOR-V rely on the PMP feature of RISC-V.
SGXIO~\cite{weiser2017sgxio} presents a hypervisor-based trusted path architecture for SGX. 
SGX-FPGA~\cite{xia2021sgx} builds a secure path between CPU and FPGA. 
To eliminate side-channel attacks,
Sanctum~\cite{costan2016sanctum} combines invasive hardware modifications with a trusted software monitor on RISC-V.

\textbf{Building TEEs with Other Hardware.}
Graviton~\cite{volos2018graviton} and StrongBox~\cite{deng2022strongbox} 
offload security-sensitive code and data to a GPU. 
Ambassy~\cite{hwang2021ambassy} and MeetGo~\cite{oh2021meetgo} use FPGA to construct TEEs, 
but they do not include softcore CPUs. 
Dedicated processor solutions, such as Google Titan~\cite{johnson2018titan}, Samsung eSE~\cite{samsunGeSE}, and Apple SEP~\cite{appleSEP}, use external connections between the REE and TEE, making them vulnerable to physical probing attacks~\cite{nasahl2020hector}.
\sysname is also inspired by other isolated execution environment solutions, including Flickr~\cite{mccune2008flicker}, TrustVisor~\cite{mccune2010trustvisor}, and Haven~\cite{haven}. 

\section{Limitations and Discussions}
\label{s:limit}

\textbf{TOCTOU.}
The presented hardware-based attestation is susceptible to TOCTOU attacks, which can be addressed by implementing SACHa~\cite{vliegen2019sacha} or RATA~\cite{de2021toctou} on top of \sysname.
SACHa presents a self-attestation framework of FPGA without a trusted hardware module,
while RATA addresses the TOCTOU attacks 
by using a hardware component to provide the context of software. 


\textbf{Low Maximum Clock Frequency and Power Consumption of FPGA.}
Even though softcore CPUs on FPGA have a low maximum clock frequency and their power consumption is always higher than hardcore CPUs with comparable performance, our experiments on a very low-end SoC FPGA device in \S\ref{s:impleval} demonstrate the practicality of \sysname.  
As the gap between FPGAs and Application Specific Integrated Circuits (ASICs) keeps reducing~\cite{kuon2007measuring} and vendors integrate FPGAs into products, e.g., AMD EPYC CPUs, it is feasible to deploy \sysname on end devices.

\textbf{Preventing Replay Attacks on Encrypted SSAs with Reference Numbers.} The current design of the SEB in SSAs is vulnerable to replay attacks. 
However, this issue can be effectively addressed by including a pair of unique reference numbers for the communicating parties. 
These reference numbers act as identifiers that help prevent attackers from intercepting and replaying the messages.

\section{Conclusion}
\label{s:conclusion}

Even though hardware-assisted TEEs have been widely adopted, they suffer from several issues that make them untrustworthy and ineffective, including static and fixed hardware TCBs and a lack of dynamic attestation of the hardware. 
In this paper, we present \sysname, a framework for building multiple TEEs on-demand with configurable hardware and software TCBs, utilizing commodity SoC FPGA devices.
\sysname establishes a dynamic root of trust that allows for full isolation and untampered execution of security-sensitive applications in enclaves from pre-existing software on the hardcore system. 
In \sysname, enclaves, which include softcore CPUs, memory, and peripherals, are created on the FPGA, and the \sysname firmware provides necessary software libraries for the applications to use. 
Additionally, \sysname offers software- and hardware-based attestation mechanisms to verify the hardware and software stacks.
We implemented \sysname on the Xilinx FPGA, and our evaluation results on the low-end Zynq-7000 system for benchmark applications and example SSAs demonstrate the effectiveness and performance of \sysname.

\section*{Acknowledgment}

We thank Paul Ratazzi, Zhenhua Wu, Kai Ni, and Arnob Paul for the discussions while this work was in progress. 
We thank Sandro Pinto for bringing TEEOD to our attention.
We also thank the organizers of MITRE eCTF 2020 for inspiring this research.
This material is based upon work supported in part by a National Science Foundation (NSF) grant (2237238), a National Centers of Academic Excellence in Cybersecurity grant (H98230-22-1-0307), and the Air Force Visiting Faculty Research Program.
Any opinions, findings, conclusions or recommendations expressed in this material are those of the author(s) and do not necessarily reflect the views of United States Government or any agency thereof.

\bibliographystyle{ieeetr}
\bibliography{reference.bib}

\appendix

\section{Experiment Device}

\begin{figure}[!h]
    \subfloat[Top view]{\includegraphics[width = \linewidth]{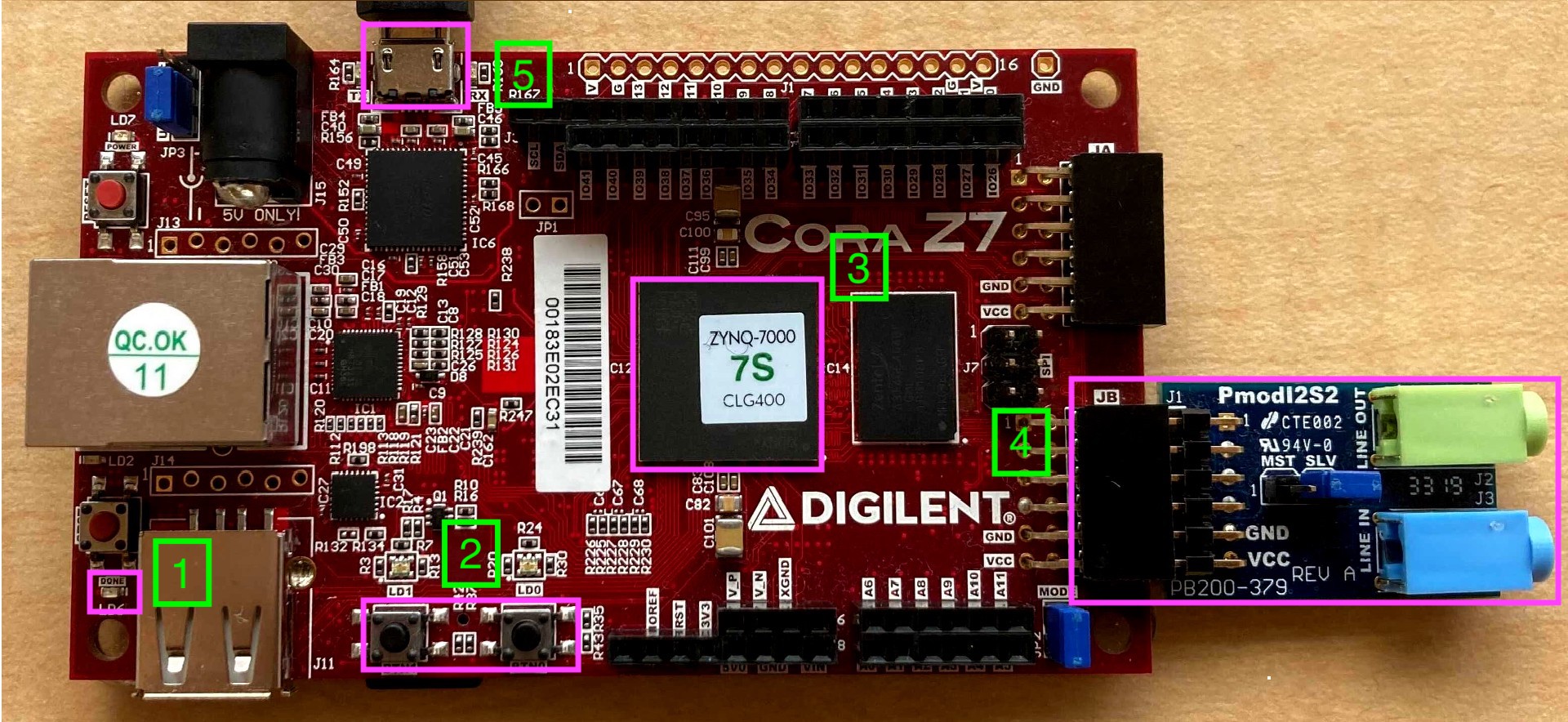}} \\ 
    \subfloat[Bottom view]{\includegraphics[width = \linewidth]{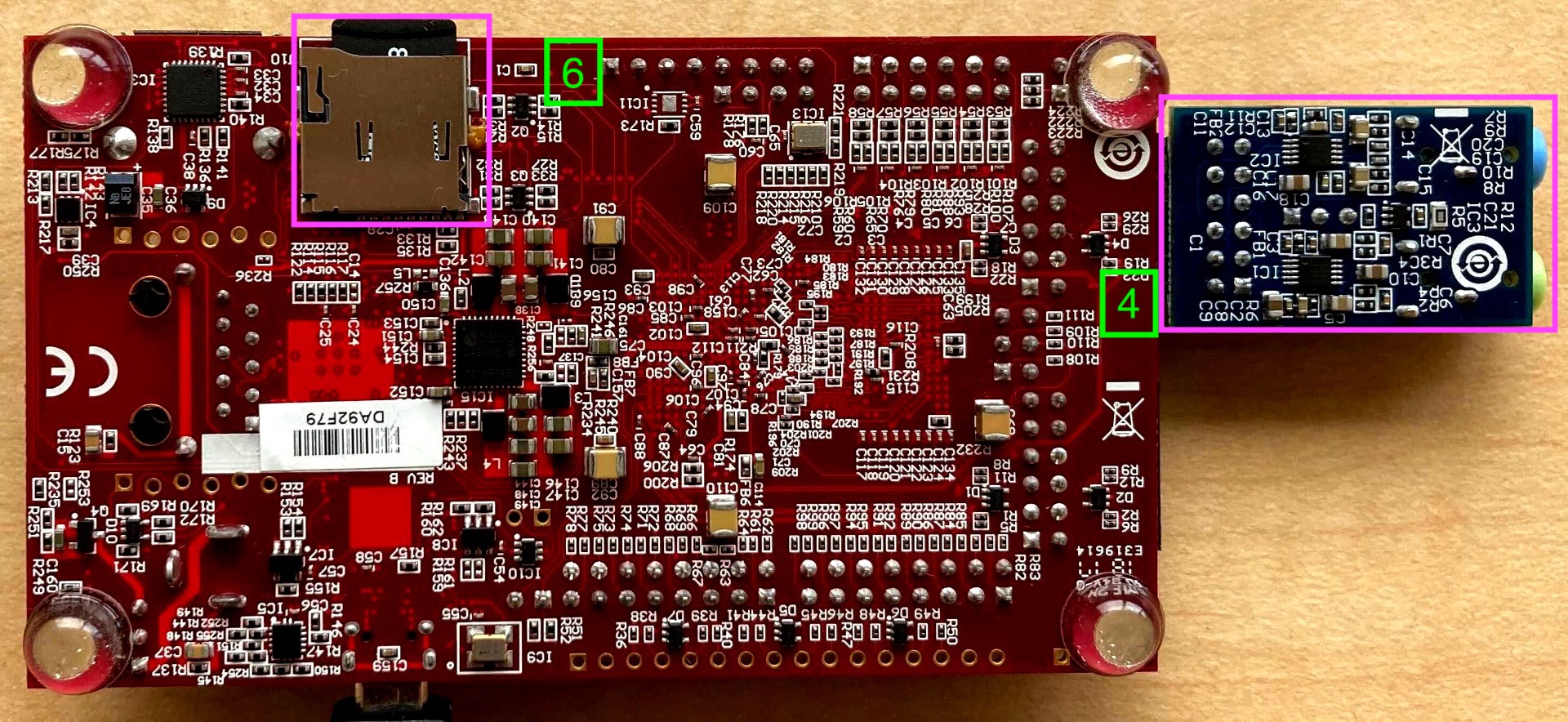}}
    \caption{The Z7-07S board with a Pmod audio module for experiments and evaluation: 
    (1) an LED as an Enclave-2 peripheral used by SSA-2,
    (2) a button as an Enclave-2 peripheral used by SSA-2, 
    (3) the Zynq-7000 SoC with a hardcore CPU and FPGA,
    (4) a Pmod port and the I2S2 stereo audio input and output device as an Enclave-3 peripheral used by SSA-3,
    (5) the on-board USB JTAG/UART for debugging and terminal output,
	(6) the SD card slot.}
    \label{fig:z7}
\end{figure}

Figure~\ref{fig:z7} shows the top and bottom view of the Cora Z7-07S development board with a single-core 667MHz Arm Cortex-A9 processor, a Xilinx Zynq-7000 FPGA, and the connected Pmod I2S2 audio device.

\section{Automatically Generated Hardware Design of the Example SSAs}

\begin{figure*}[!t]
    \centering
    \subfloat[Hardware for SSA-1 with a Debugging Module]{\includegraphics[width = 0.49\linewidth]{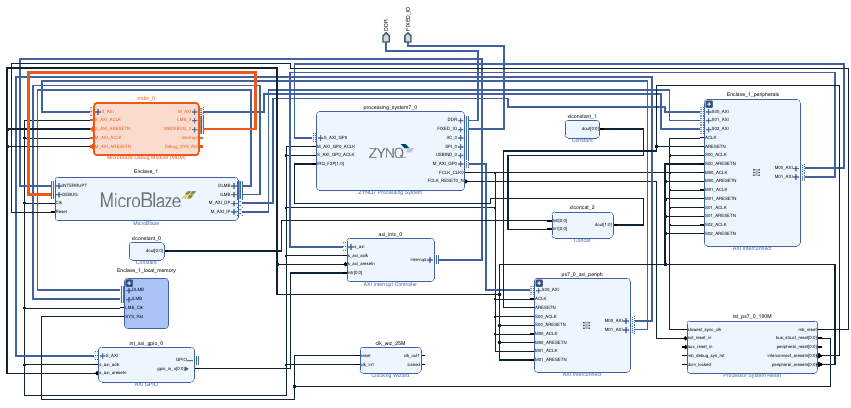}} \hfill
    \subfloat[Hardware for SSA-1 without a Debugging Module]{\includegraphics[width = 0.49\linewidth]{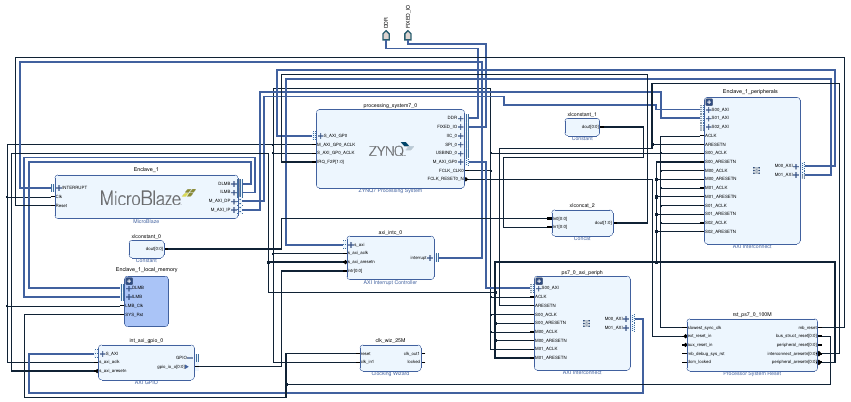}}\\
    \subfloat[Hardware for SSA-2 without a Debugging Module]{\includegraphics[width = 0.49\linewidth]{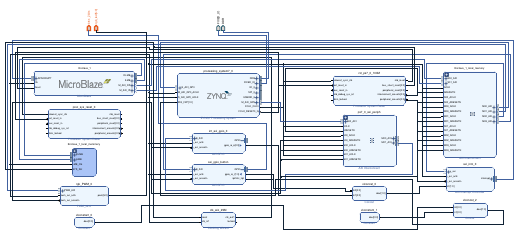}} \hfill
    \subfloat[Hardware for SSA-3 without a Debugging Module]{\includegraphics[width = 0.49\linewidth]{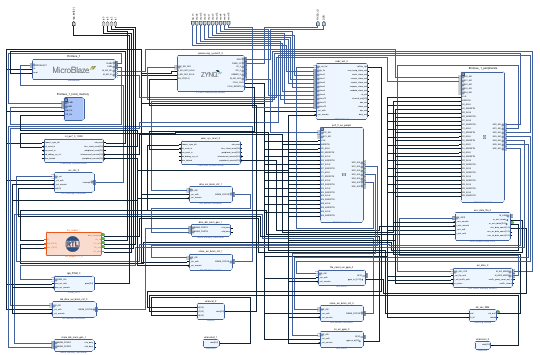}}
    \subfloat[Hardware for SSA-4 without a Debugging Module]{\includegraphics[width = 0.49\linewidth]{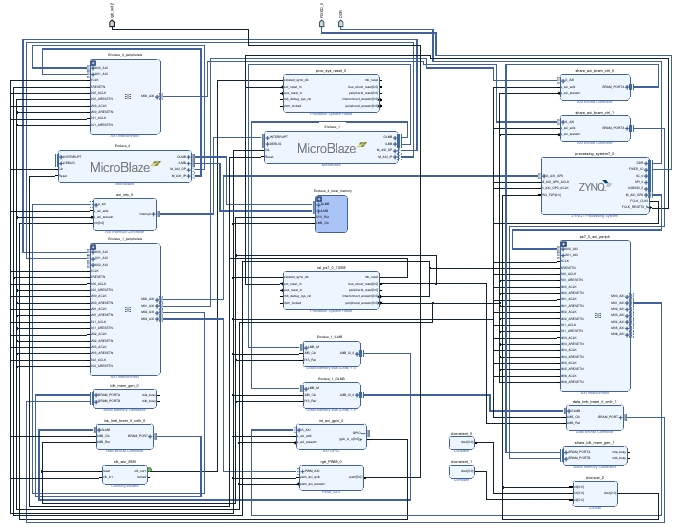}}    
    \cprotect\caption{Block diagrams of the hardware designs of the four enclaves for the four example SSAs (software-based attestation profile). The ZYNQ processing system (\verb|processing_system7_0|) represents the hardcore Cortex-A processor.  Each of the enclaves for SSA-1, SSA-2, and SSA-3 has one MicroBlaze softcore CPU, whereas SSA-4 has two enclaves. 
The MicroBlaze interrupt interface is connected to an AXI interrupt controller and configured with an AXI GPIO.
All output GPIO to the softcore is connected to the hardcore system for triggering interrupts.
Two AXI Interconnects, \verb|mb_axi_mem_interconnect_0| for Microblaze and \verb|ps7_axi_periph| for Cortex-A processor are used to connect external IPs.
The ZYNQ, Microblaze, and other IPs get primary clock input from the \verb|clk_in1| port.
The reset port is connected to the reset interfaces of the IPs.}
    \label{fig:corr-res}
\end{figure*}

Figure~\ref{fig:corr-res}(a) shows the hardware for SSA-1 with a MicroBlaze Debugging Module (MDM)~\cite{mdmTRM} to help the developers debug SSAs.
A master interface of MDM is also connected to the \verb|ps7_axi_periph| to enable debugging from the hardcore system side.
Figure~\ref{fig:corr-res}(b) shows the same hardware configuration without the debugging module.
As shown in Figure~\ref{fig:corr-res}(c), the hardware design of SSA-2 includes two peripherals that are only connected to the FPGA.
A pulse width modulation IP~\cite{pwm} connects the \verb|RGB_LED| ports, and an AXI GPIO IP~\cite{axiGpio} connects the button ports to the \verb|mb_axi_interconnect_0| interconnect.
As shown in Figure~\ref{fig:corr-res}(d),
the hardware for SSA-3 includes several more IPs for different functionality.
An I2S transmitter RTL (SPI) is added to interact with the Pmod I2S2 module Codec module.
Additionally, an AXI direct memory access IP~\cite{axiDMA} with BRAM and an AXI stream data FIFO IP~\cite{axififo} to offload audio data streaming computation from Microblaze are inserted by the \hbuilder.
An ADC module for reading voltages, which is used for performance monitoring~\cite{xadc}, is also connected.
Figure~\ref{fig:corr-res}(e) represents the hardware of SSA-4 with two enclaves.
Both the enclaves have their own BRAM, 128KB, and 32KB respectively.
Enclave-1 has a shared DRAM with Zynq processor as SEB for communication, while Enclave-2 does not have any SEB with Zynq.
Instead of inter enclave communication, one 8KB BRAM is shared between Enclave-1 and Enclave-2 without the Zynq processing system access.

\end{document}